\documentstyle[epsfig]{mn}

\newcommand{\ed}{{\rm{d}}}

\title[Baryon density fluctuations in cosmological simulations]
{The evolution of baryon density fluctuations
in multi-component cosmological simulations}
\author[N. Yoshida, N. Sugiyama \& L. Hernquist]
{Naoki Yoshida$^{1,2}$\thanks{E-mail:
nyoshida@cfa.harvard.edu (NY)},
Naoshi Sugiyama$^{2}$, Lars Hernquist$^{1}$\\
$^1$Harvard-Smithsonian Center for Astrophysics, 60 Garden Street, Cambridge MA 02138, USA\\
$^2$National Astronomical Observatory Japan, Mitaka, Tokyo 181-8588, Japan}
\date{Accepted for publication in MNRAS, May 20, 2003}

\begin{document}
\maketitle

\begin{abstract}

We critically examine how the evolution of the matter density field in
cosmological simulations is affected by details of setting up initial
conditions.  We show that it is non-trivial to realise an initial
distribution of matter in $N$-body/hydrodynamic simulations so that
the baryon and dark matter density fluctuations and their velocities
evolve consistently as theoretically predicted.  We perform a set of
cosmological simulations and use them to distinguish and verify an
appropriate method for generating initial conditions.  We show that a
straightforward way of applying the Zel'dovich approximation to each
component using distinct transfer functions results in an incorrect
growth of density fluctuations and that it is necessary to correct
velocities at the initial epoch.  The unperturbed uniform particle
distribution must be also generated appropriately to avoid tight
coupling of the baryonic and dark matter components.  We recommend
using independent ``glass'' particle distributions, using distinct
transfer functions for baryons and dark matter, {\it and} taking into
account the difference in the velocity fields at the initialisation
epoch.  The proposed method will be useful for studies of the
evolution of the intergalactic medium and the formation of the first
cosmological objects using numerical simulations.

\end{abstract}


\section{Introduction}

High-precision measurements of the Cosmic Microwave Background (CMB)
radiation by the {\it WMAP} satellite provide strong support for the
so-called standard model of cosmic structure formation, in which Cold
Dark Matter (CDM) and dark energy dominate the Universe.  The matter
distribution at the decoupling epoch has been directly probed by
measurements of CMB anisotropies, and it has been confirmed that the
observed matter distribution is consistent with the predictions of
popular inflationary theories (Spergel et al. 2003).  Within this
theoretical framework, structure formation in the early Universe is
described by gravitational amplification of small initial density
fluctuations; non-linear evolution leads to the formation of dark
matter halos, followed by hydrodynamic processes such as gas infall
into gravitational potential wells, shock heating, and radiative
cooling (e.g. Yoshida et al. 2003).  Precise modelling of the angular
power spectrum of the CMB anisotropies determines the relative
densities of the baryonic and non-baryonic components.  Furthermore,
accurate solutions of the nonlinear Boltzmann equations (Hu \&
Sugiyama 1995; Ma \& Bertschinger 1995; Seljak \& Zaldarriaga 1996)
show that there is a substantial difference in the distribution of
baryons and dark matter at the decoupling epoch.  Indeed, recent
detailed numerical studies (Yamamoto, Sugiyama \& Sato 1998; Singh \&
Ma 2002) show that the differences remain until much lower redshift $z
\sim 20$ for high baryon densities. In fact, the ratio of baryon density
to CDM density determined by the {\it WMAP} data (Spergel et al. 2003)
is as large as $\Omega_{\rm baryon}/\Omega_{\rm cdm} \sim 0.2$.
Thus these results clearly contradicts the assumption, often made in the
context of structure formation, that baryons trace the dark matter;
i.e. that local baryon densities are just proportional to those of the
dark matter, and their velocities are identical both in direction and
in amplitude.

Numerical simulation of structure formation is aimed at revealing how
the large-scale features of the Universe formed and evolved from high
to low redshifts.  Conventionally, cosmological $N$-body simulations
are started from an initial matter distribution -- a random Gaussian
realisation -- generated using an input power spectrum for the {\it
total} matter density.  Although this is obviously the most plausible
way of generating initial conditions for $N$-body simulations with
only a single collisionless component, it is non-trivial to realise
the matter distribution in simulations that employ two (or more)
components so that the density and velocity fields of both components
evolve consistently.  A common belief is that details in the initial
configuration are erased by non-linear evolution and hence do not
matter if one is interested in the structure of non-linear objects
(i.e. ``dark halos'') at low redshift.  However, this reasoning may
not apply at high redshifts or for the intergalactic medium (IGM)
where the density fluctuations on relevant scales are in the linear to
mildly non-linear regime.

In numerical studies of the properties of the IGM, such as those
concerned with the reionisation history of the Universe (e.g. Gnedin
\& Ostriker 1997; Gnedin 2000; Sokasian et al. 2002, 2003), it is
important to accurately compute the thermal properties of baryons in
low density regions.  Various analytic methods have been developed to
obtain the local densities and temperatures of the IGM from the
distribution of dark matter that can be computed relatively accurately
by either direct $N$-body simulations or linear theory (e.g. Bi, B\"orner \& Chu
1992; Jones 1996; Hui \& Gnedin 1997; Nusser 2000; Matarrese \& Mohayaee 2002). On the other
hand, in cosmological $N$-body/hydrodynamic simulations, simplifying
assumptions are often made. For example, in cosmological Smoothed
Particle Hydrodynamics (SPH) simulations, gas particles are either put
on top of dark matter particles, or simply displaced by half the mean
interparticle separation before perturbations are imposed (e.g. Katz
et al. 1996).  A naive way of improving upon this simple approach
would be to apply the Zel'dovich approximation to baryonic and dark
matter components separately using separate transfer functions for
each.  It is unclear, however, if the generated initial particle distributions
and the velocity field reproduce consistent results with predictions
of full non-linear theory.

Another closely related issue is that, in particle simulations, a
smooth density field must be represented by discrete mass elements. It
is well-known that discretisation itself imposes some limitations on
the accuracy of numerical simulations (Splinter et al. 1998; Hamana,
Yoshida \& Suto 2001; Baertschiger et al. 2002).  Baertschiger \&
Sylos-Labini (2001) show that the discreteness of the particles
contributes to the evolution of power-law clustering on small scales.
G\"{o}tz \& Sommer-Larsen (2002) argue that even small differences in
unperturbed particle distributions can yield quite different, in some
cases unfavourable, results in the nonlinear evolution of systems (see
their Fig. 1).

These facts and questions motivate the need for a systematic study in
order to develop and verify a proper method for setting up initial
conditions for multi-component cosmological simulations.  In the
present paper, we examine how details in the initial set-up affect the
evolution of matter density fluctuations.  We specifically study the
power spectra of baryonic and dark matter components and the abundance
of dark matter halos using a set of $N$-body simulations.  We show
that differences in initial configurations indeed result in
substantial differences in the evolved density field and its growth
rate.  We compare the numerical results with theoretical predictions
and determine an improved method for generating initial conditions.

The rest of the paper is organised as follows.  In Section 2, we
present the basic theory of the evolution of density fluctuations in
the early Universe.  In Section 3, we describe the simulation set.  We
show our numerical results in section 4.  Discussion is given in
Section 5.

\begin{figure}
\centering
\epsfxsize=\hsize\epsffile{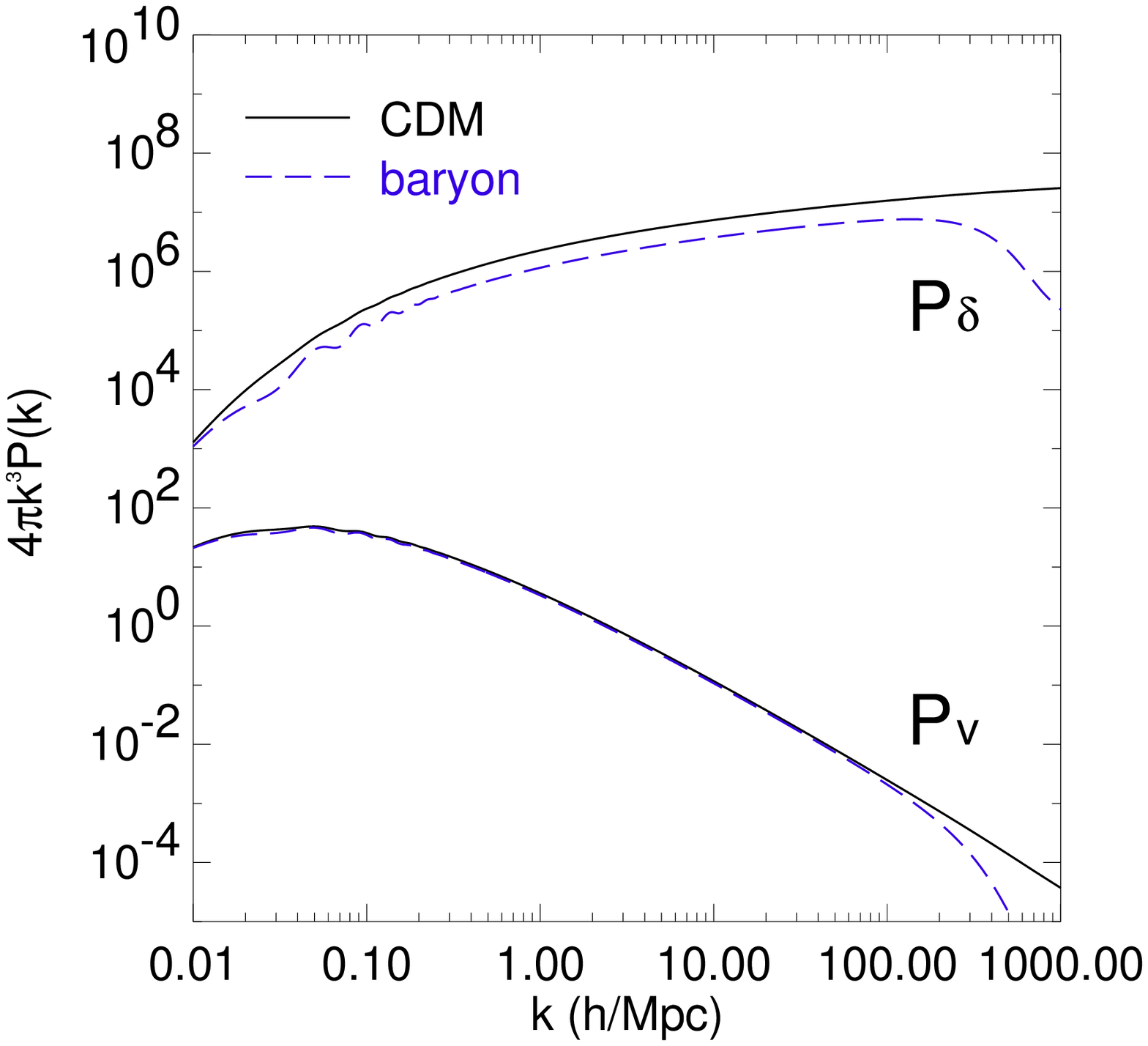}\\
\epsfxsize=\hsize\epsffile{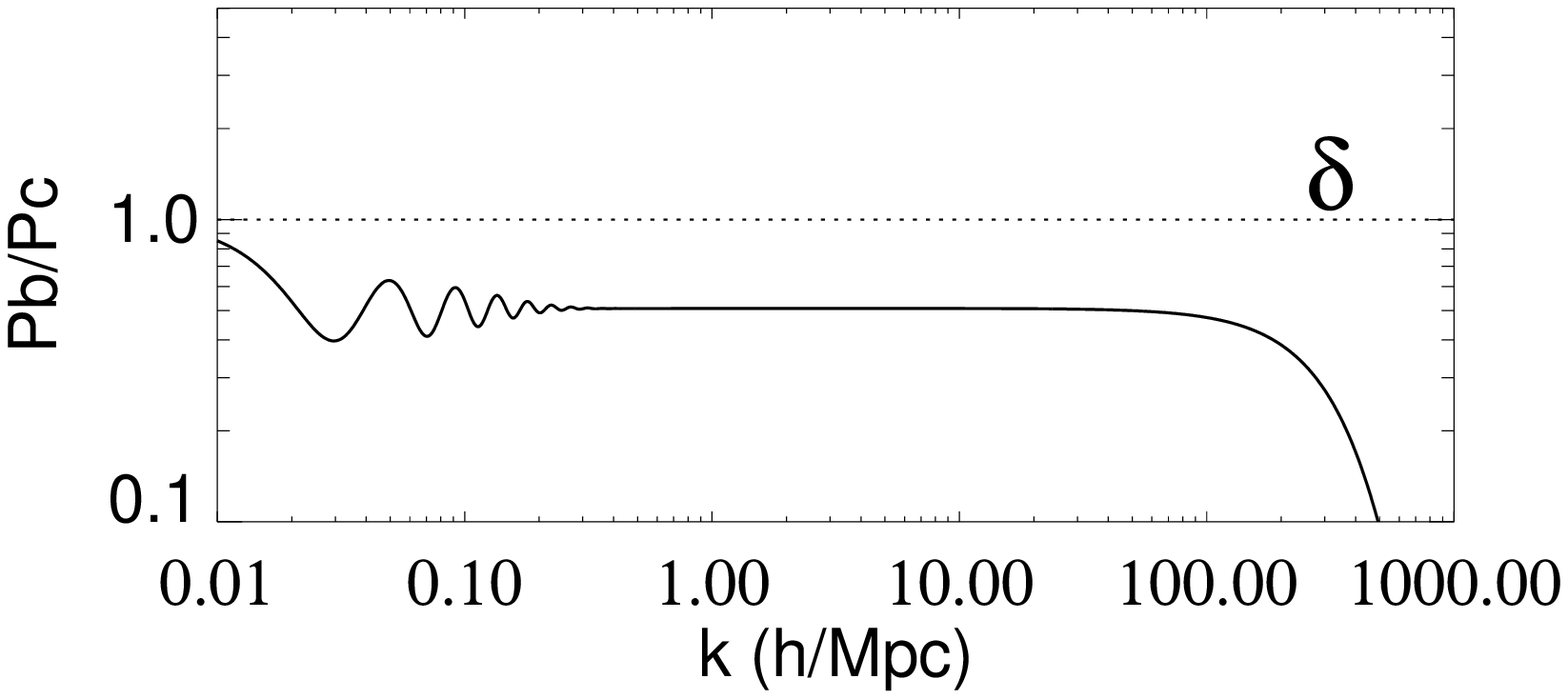}\\
\epsfxsize=\hsize\epsffile{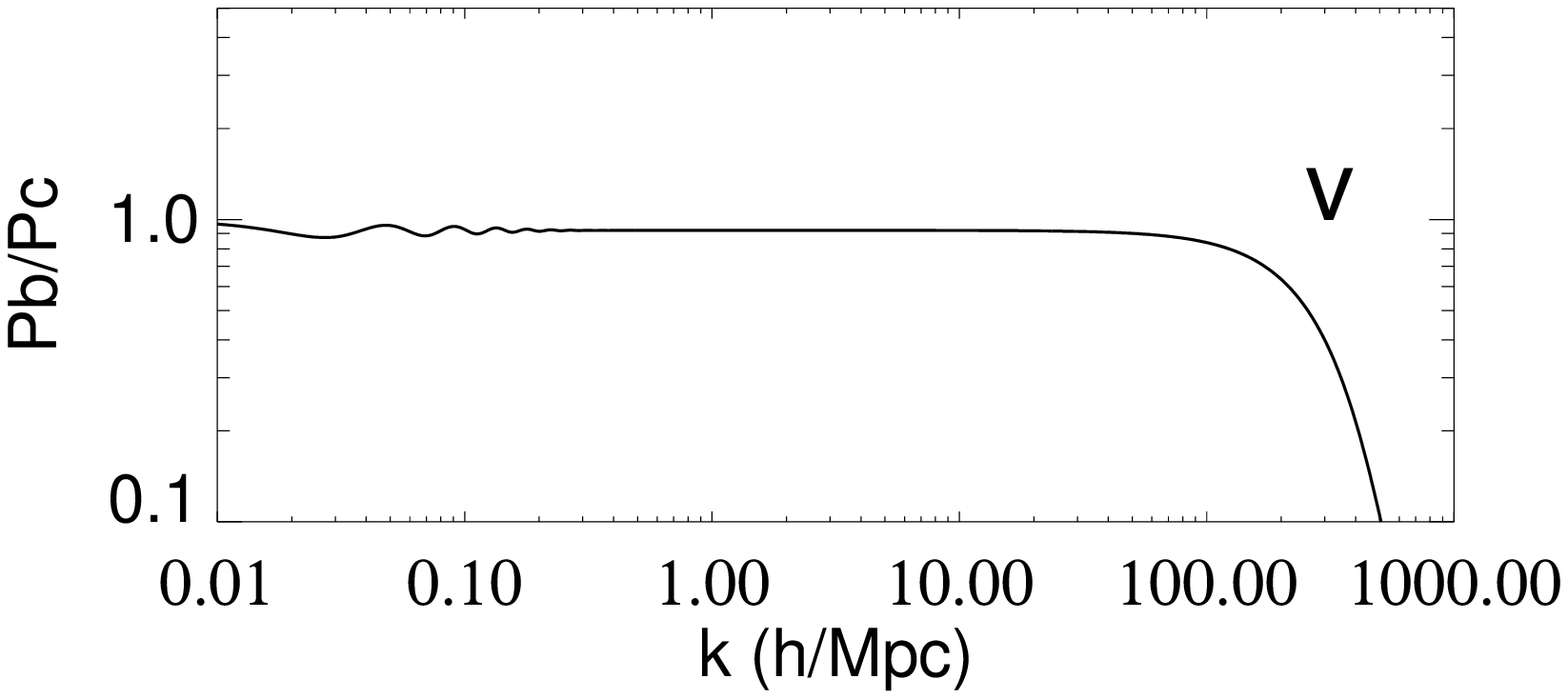}
\caption[Power spectra]
{The density and velocity power spectra for CDM (solid line) and for
baryons (dashed line) at $z=100$ for a standard $\Lambda$CDM model
with $\Omega_{\rm cdm}=0.26$, $\Omega_{\rm baryon}=0.04, \Omega_{\Lambda}=0.7,
H_{0}=70$ km s$^{-1}$Mpc$^{-1}$. 
In the top panel, the relative amplitudes of the density to velocity power spectra are scaled
arbitrarily. The middle panel shows the ratio of the two density power
spectra and the bottom panel shows that of the two velocity power
spectra.}
\label{fig1}
\end{figure}

\section{Theory}

\subsection{Evolution of baryon density fluctuations}

The evolution of the cosmological baryon perturbations before the
decoupling epoch is described by a number of physical processes which
couple the perturbation fields of baryons, CDM, photons and neutrinos
in a complex manner.  One thus needs to solve the full set of coupled
nonlinear Einstein-Boltzmann equations in order to obtain an accurate
result.  Recent detailed studies by Liu et al. (2001) and Singh \& Ma
(2002) show that, although density fluctuations on scales smaller than
the photon diffusion scale (Silk 1968) are strongly suppressed, the
growth of the baryon density fluctuations are eventually accelerated
{\it during} the drag epoch because the photon-baryon coupling breaks
down by recombination.  It is also important to note that second-order
effects arising from the coupling of the velocity fields to the baryon
density field can enhance the growth of small scale fluctuations
(Shaviv 1998; Liu et al. 2001; Singh \& Ma 2002), although this effect
plays an important role only on very small length scales ($k \gg 1000$Mpc$^{-1}$).

After recombination, gravitational infall causes the baryon density
fluctuations to catch up with those of the dark matter.  Yamamoto et
al. (1998) show, however, that the growth of perturbation modes on
length scales smaller than the Jeans length is delayed due to the
initial oscillatory behaviour of the baryons, and thus the residual
differences in the fluctuation amplitudes between baryons and dark
matter remain until redshifts $z \sim 20 - 100$.  This is clearly
shown in Figure 1.  There we plot the density power spectra and the
velocity power spectra for baryons and CDM for a conventional
$\Lambda$CDM universe (parameters as summarised in Section 3) at
$z=100$.  The power spectra are computed using the Boltzmann solver of
Sugiyama (1995), as described in Yamamoto, Sugiyama \& Sato (1998).
The density power spectra for baryons and CDM differ by about a factor
of two for $1 < k < 10$, whereas the difference in the velocity power
spectra is less than 10 percent on these scales (compare the middle
panel with the bottom panel).

\subsection{The Zel'dovich approximation}

The standard procedure for setting up initial conditions for
cosmological $N$-body simulations is to perturb a uniform particle
distribution using the Zel'dovich approximation (Zel'dovich 1970).  An
extensive study of the applicability of the Zel'dovich approximation
is presented in a series of papers by Valageas (2002a,b and references
therein).  We summarise the elements of the method in this section.
 
Let the initial Lagrangian coordinate of a particle in an unperturbed
distribution be ${\bf q}$. Then each particle is subject to a
displacement corresponding to a density perturbation. In the
Zel'dovich approximation the Eulerian coordinate of the particle at
time $t$ is
\begin{equation}
{\bf r}(t,{\bf q})=a(t)[{\bf q}-b(t)\nabla_{\bf q}\Phi_{0}({\bf q})],
\label{eq_zeldovich}
\end{equation}
where ${\bf r}=a(t){\bf x}$ with ${\bf x}$ being a comoving coordinate
and $b(t)$ is the growth factor.  The velocity field is obtained from
\begin{equation} 
{\bf v}=a\frac{\ed{\bf x}}{\ed t}=-a\dot{b}\nabla_{\rm q}\Phi_{0}({\bf q}).
\label{eq_za_v}
\end{equation}
The quantity $\Phi_{0}({\bf q})$ is related to the density perturbation 
$\delta$ in the linear regime via the Poisson equation:
\begin{equation}
\delta = b \nabla^{2}_{\rm q}\Phi_{0}.
\end{equation}
It is important to point out that the initial velocity field and the
displacement field are linearly related via equation (\ref{eq_za_v}).
Namely, particles in the Zel'dovich approximation execute motion on
straight line trajectories.  Therefore, if one uses different transfer
functions for two components (e.g. CDM and baryons) while keeping the
phase information for a random Gaussian field identical, the generated
displacement fields and hence the velocity fields for the two
components should differ by an amount corresponding to the difference
in the amplitudes of the transfer functions.  However, after the
decoupling epoch, baryons fall into gravitational potential wells that
are dominated by the dark matter, whereas the growth of dark matter
density fluctuations is slightly delayed due to the baryonic matter
which is more homogeneously distributed.  Then, the difference in
velocities of baryons and dark matter decays quickly after the
decoupling epoch (compare the middle panel and the bottom panel in
Figure 1). 

 Nusser (2000) and Matarrese \& Mohayaee (2002) recently derived 
analytic solutions for the evolution of linear density density perturbations
for both baryons and dark matter.
While their solutions fully describe the
evolution of linear density fluctuations of a two-component fluid for a few 
particular cases, they are not directly applicable to generating initial
conditions for direct $N$-body simulations in more general cases.
In the present paper we develop a more practical method.
The key question is {\it how can we generate
initial conditions using solutions of a full non-linear calculation ?}

We test five different methods for setting up initial conditions in
order to distinguish a correct method (if one exists) and to quantify
its accuracy.  We describes the technical details in the next section.

\begin{table}
\begin{center}
\caption{Simulation sets}
\begin{tabular}{ccccc}
\hline
 --   & glass files & trans. func. & phase & velocity correction \\
Run A & same & diff         & same & no \\
Run B & diff & diff         & same & no \\
Run C & diff & diff         & same & yes \\
Run D & same & same
\footnote{Transfer function for the total matter} & same & no \\
Run E & diff & diff         & diff & no \\
\hline
\end{tabular}
\end{center}
1 Transfer function for the total matter
\end{table}

\section{The $N$-body/SPH simulations}

We use the parallel $N$-body/SPH code GADGET (Springel, Yoshida \&
White 2001), in its ``conservative entropy'' version (Springel \&
Hernquist 2002, 2003).  All our simulations employ $2\times 128^3$ particles
(CDM and non-radiative gas components) in a cosmological box of 4
$h^{-1}$Mpc on a side.  We work with a flat $\Lambda$-dominated Cold
Dark Matter universe, with matter density $\Omega_{\rm cdm}=0.26$,
$\Omega_{\rm b}=0.04$, cosmological constant $\Omega_{\Lambda}=0.7$
and expansion rate at the present epoch $H_{0}=70$km s$^{-1}$Mpc$^{-1}$.
The qualitative features of our results are not affected by the
adopted cosmology, although the degree of the numerical artifacts we
discuss depend on the ratio of $\Omega_{\rm cdm}/\Omega_{\rm b}$.
Table 1 summarises the five methods we consider for the set-up of the
initial conditions. We describe each specifically in the following
subsections.  We note that for all the simulations we set the
parameters governing the accuracy of the force calculation and time
stepping to be very conservative, following Springel et al. (2001) and
Power et al. (2003). 
We set the gravitational softening length $\epsilon_{\rm soft}$
= 3 $h^{-1}$kpc $\sim$ one tenth of the mean inter-particle separation.

\subsection{Unperturbed particle distribution}

There are two popular methods for distributing particles
``uniformly''.  One is to place particles on a rectangular grid.
While used most often, a grid distribution has the unfavourable
feature of imposing a particular direction in the particle
distribution.  In other words, a grid distribution is not isotropic,
although it is homogeneous on scales larger than the mean
inter-particle separation (see e.g. Baertschiger \& Sylos-Labini 2001).  
G\"{o}tz \& Sommer-Larsen (2002) report
that grid distributions cause peculiar problems when used for warm
dark matter simulations such as those of Bode, Ostriker \& Turok
(2001).  
Although the degree of these
problems is somewhat uncertain and is likely to be case dependent, it
is preferable to avoid these issues altogether by using ``glass''
particle distributions.  Following White (1994), we generate glass
initial conditions by evolving a gravitationally interacting
$N$-particle system in an expanding background with the sign of
gravity reversed such that each particle feels repulsive forces from
all the other particles.  When the system reaches a quasi-equilibrium
state, unphysical clumping in the initial distribution is
substantially damped, while maintaining uniformity. The particle
distribution is also nearly isotropic and hence does not possess any
preferred direction.

For all the numerical experiments in the present paper, we use glass
particle distributions.  We generate two independent glass
distributions and use them for gas and CDM particles in Runs B, C, and
E, whereas we use identical glass particle distributions for both
components in Run A and D (see Table 1).

\subsection{Transfer functions}

In the context of structure formation in CDM models, the density power
spectrum at a given epoch is related to the transfer function by
\begin{equation}
P(k)=A k^n T^2(k),
\end{equation}
where $A$ is the normalization factor.
We consider the Harrison-Zel'dovich primordial power spectrum with
$n=1$. Figure 1 shows the power spectrum $4 \pi k^3 P(k)$ at $z_{\rm
init}=100$.  On very large length scales ($k < 0.01 h$/Mpc), the power
spectra for baryons and CDM have the same amplitudes, whereas on
intermediate length scales ($1 < k < 100$) the amplitudes for baryon
density perturbations are smaller by about a factor of two at $z=100$.
Also the power spectrum for baryons shows characteristic wiggles at
$0.01 < k < 0.2$ which is induced by Jeans oscillations before the recombination
epoch.
  
For Runs A, B, and C, transfer functions are given separately for the
CDM and the baryonic components, whereas for Run D, we use a single
transfer function computed for the total matter density
fluctuations.  The matter transfer function for Run D is computed from
\begin{equation}
T(k)=\frac{\Omega_{\rm b}}{\Omega_{\rm total}}T_{\rm b}(k) 
+\frac{\Omega_{\rm cdm}}{\Omega_{\rm total}}T_{\rm cdm}(k). 
\end{equation}

\subsection{Phases}

In standard inflationary theories that predict adiabatic-type density
fluctuations, the distribution of various forms of matter is described
by essentially a single scalar function, and thus the phases for
baryons and CDM are identical for all the perturbation modes.
Although we have no compelling reason to assume the phases are
different on any length scale, it is interesting to examine the effect
of varying the phases. In order to make comparisons, we assign random,
independent phase information for baryons and CDM in Run E.  Other
configurations for Run E are identical to those of Run B.  We compare
these simulations in Section 5.

\subsection{Velocity power spectrum}

As we have discussed in Section 2, the differences in the velocity
power spectra of baryons and CDM are smaller than the differences in
fractional amplitudes of density fluctuations (Figure 1).  Since the
fractional difference is within 10\% and nearly a constant at $z_{\rm
init}=100$ over most of the relevant scales, we approximate the
velocity fields of both CDM and baryons by the velocity field of the
{\it total} matter.  In practice, we compute the velocity field using
the transfer function for the total matter while keeping the phase
information identical to that used in generating the spatial
displacements.  Note that this is equivalent to multiplying the
velocity term in equation (\ref{eq_za_v}) by a scalar factor. The
scalar factor could, in principle, be made scale-dependent, as is done
by Klypin et al. (1997) for Mixed Dark Matter simulations.

\begin{figure*}
\epsfxsize=0.45\hsize\epsffile{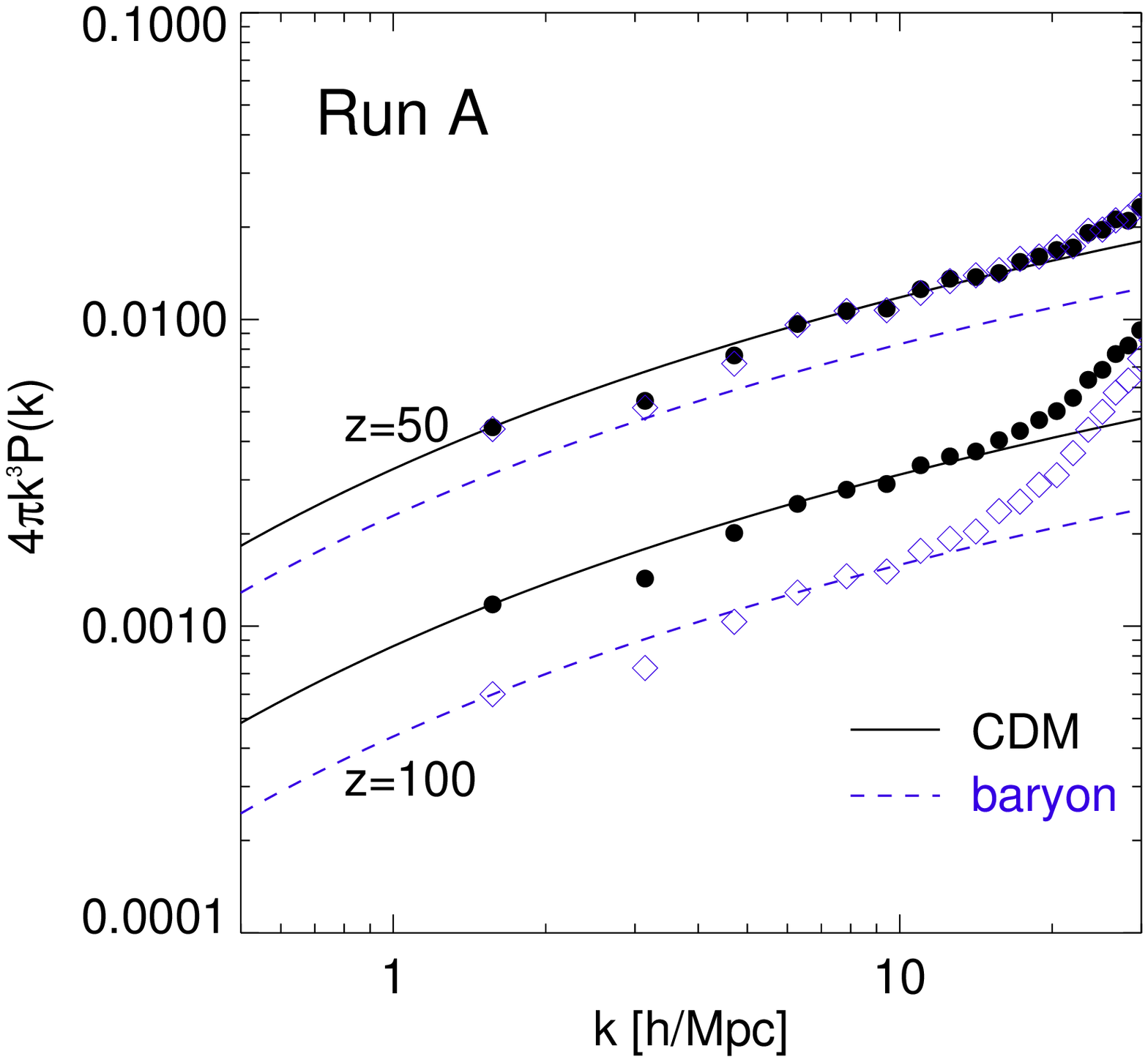}%
\epsfxsize=0.45\hsize\epsffile{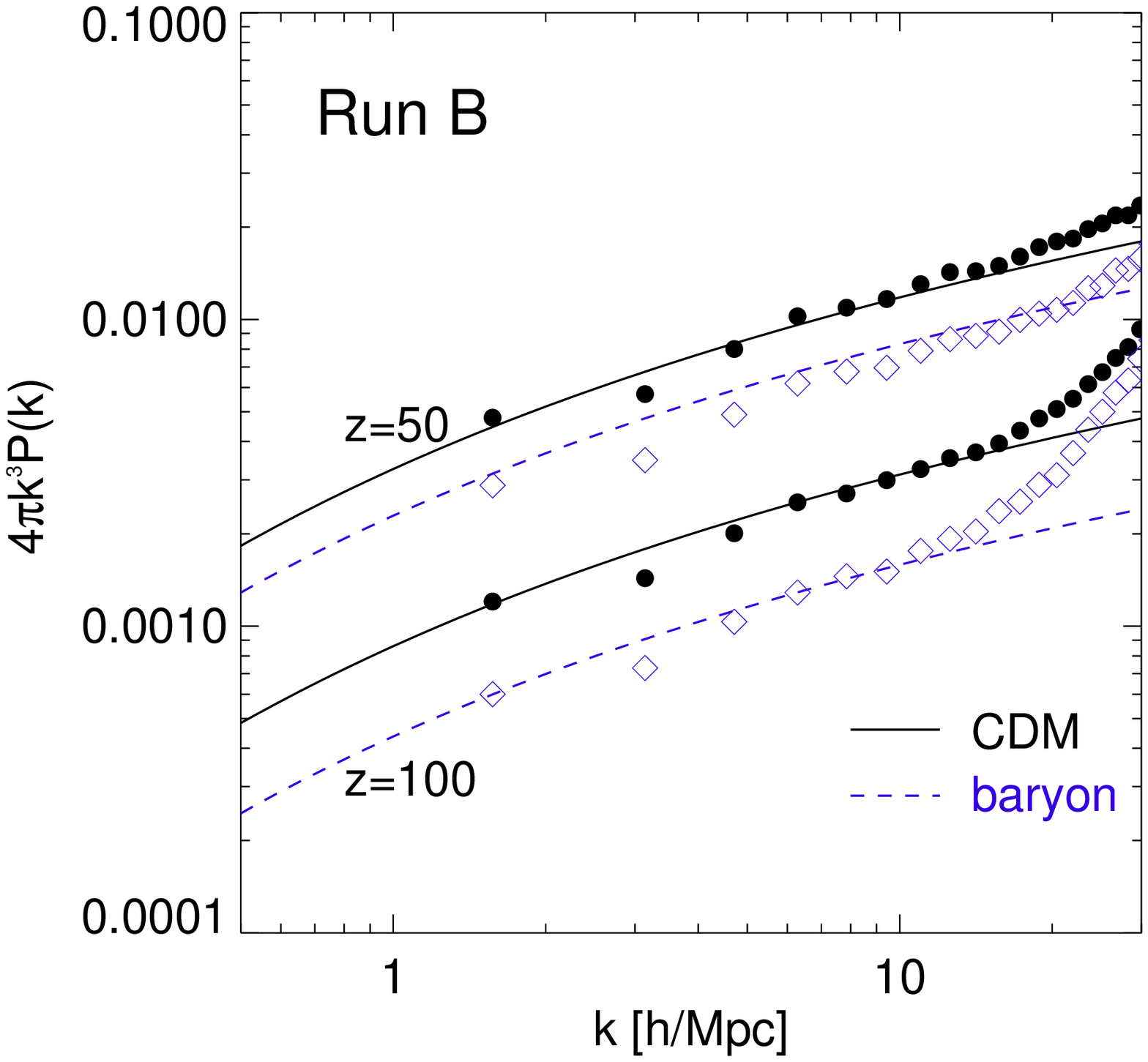}\\ 
\epsfxsize=0.45\hsize\epsffile{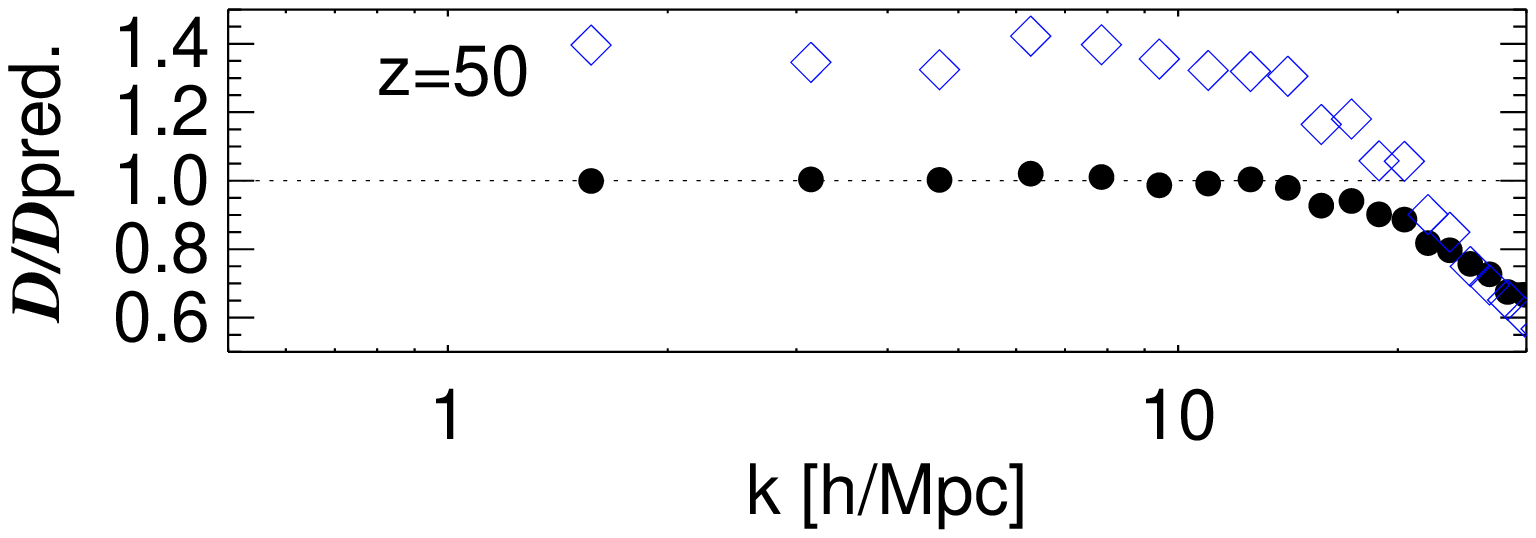}%
\epsfxsize=0.45\hsize\epsffile{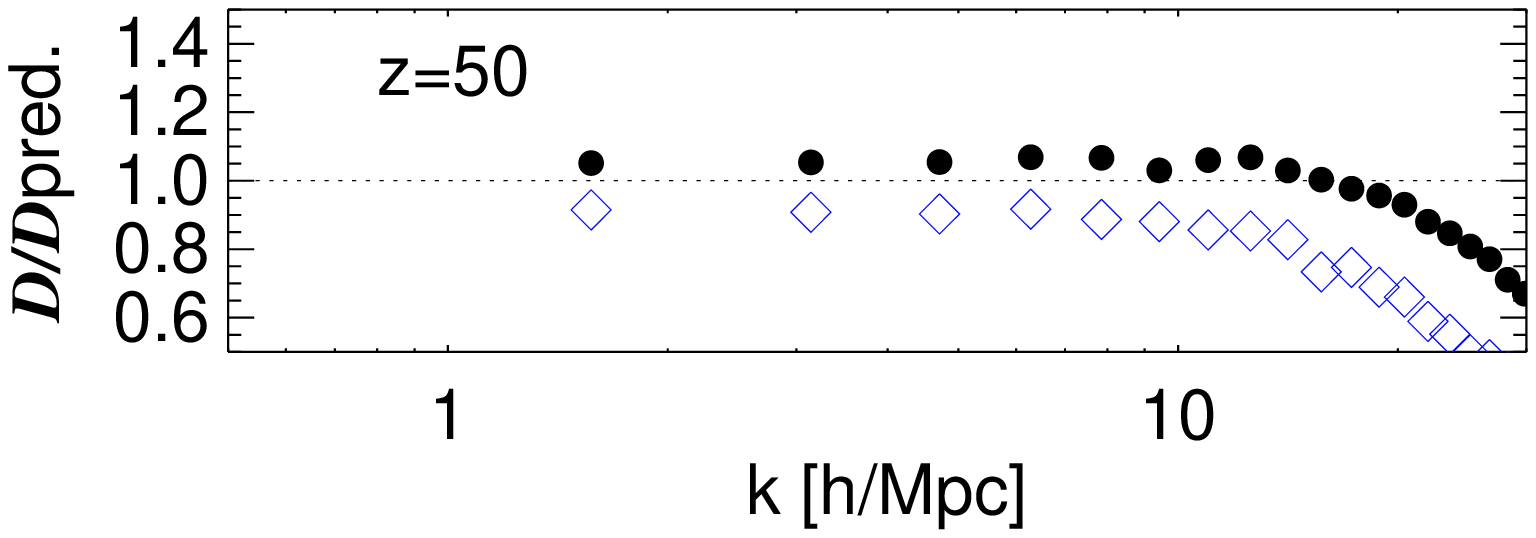}\\  
\epsfxsize=0.45\hsize\epsffile{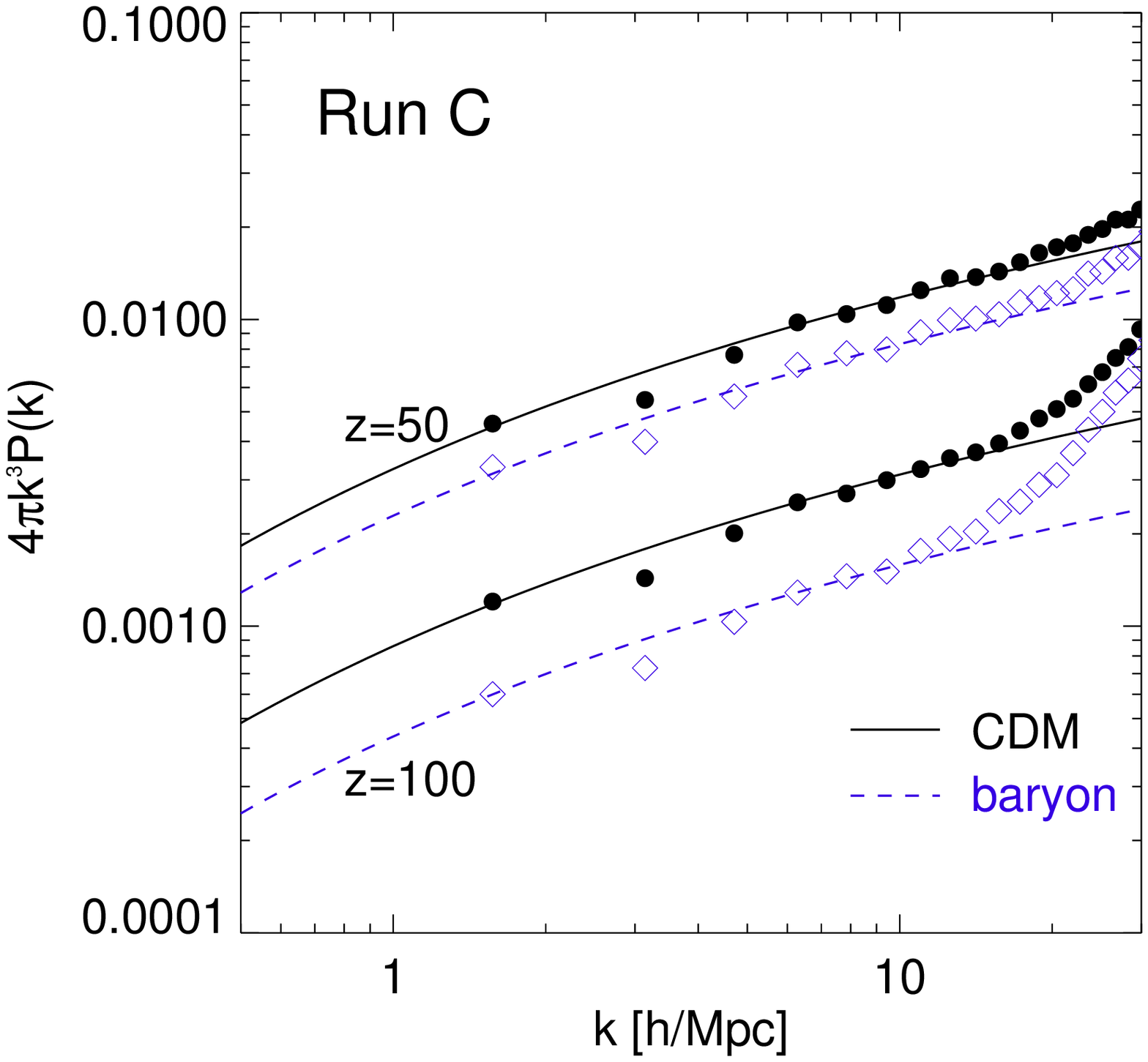}%
\epsfxsize=0.45\hsize\epsffile{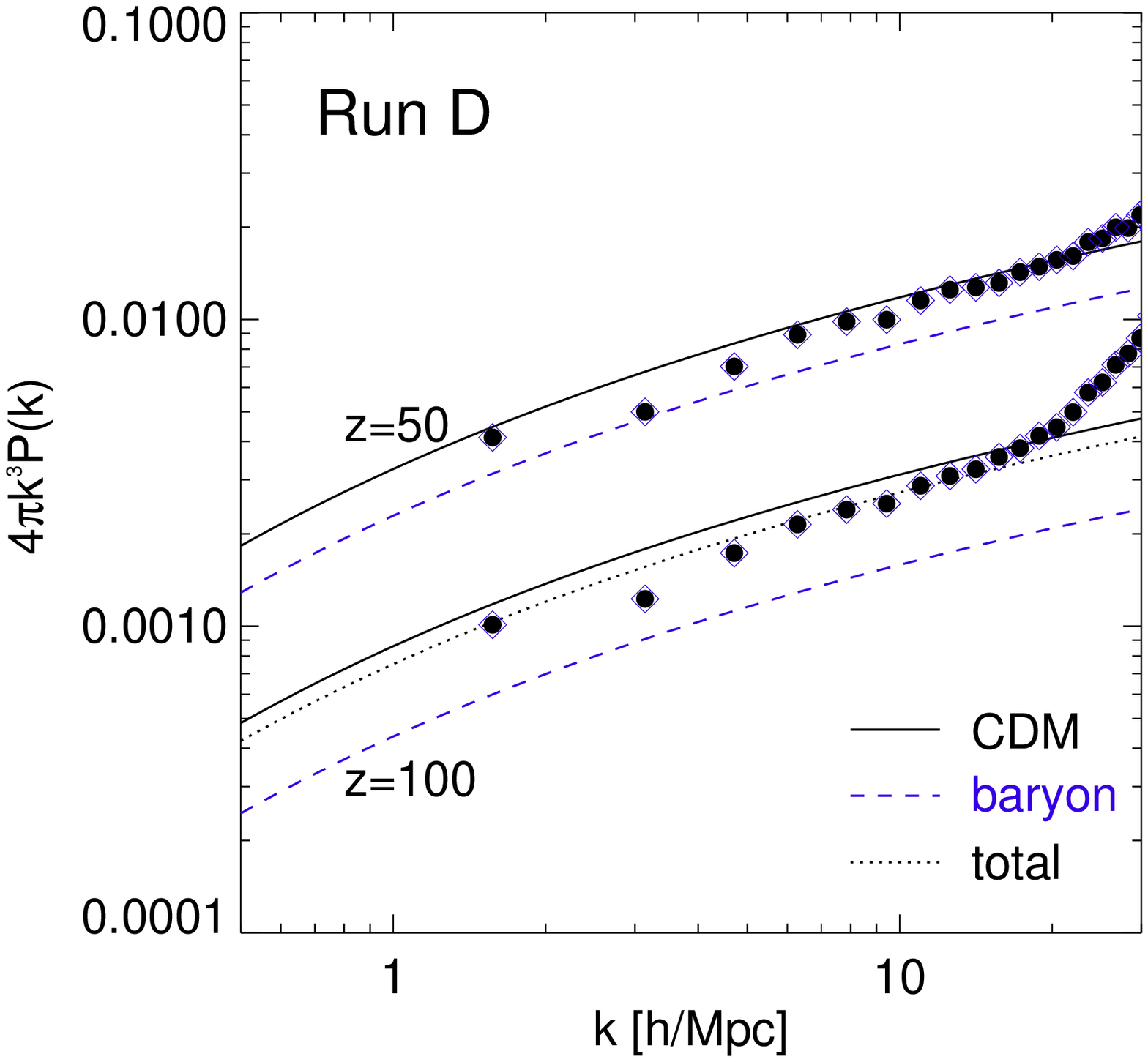}\\  
\epsfxsize=0.45\hsize\epsffile{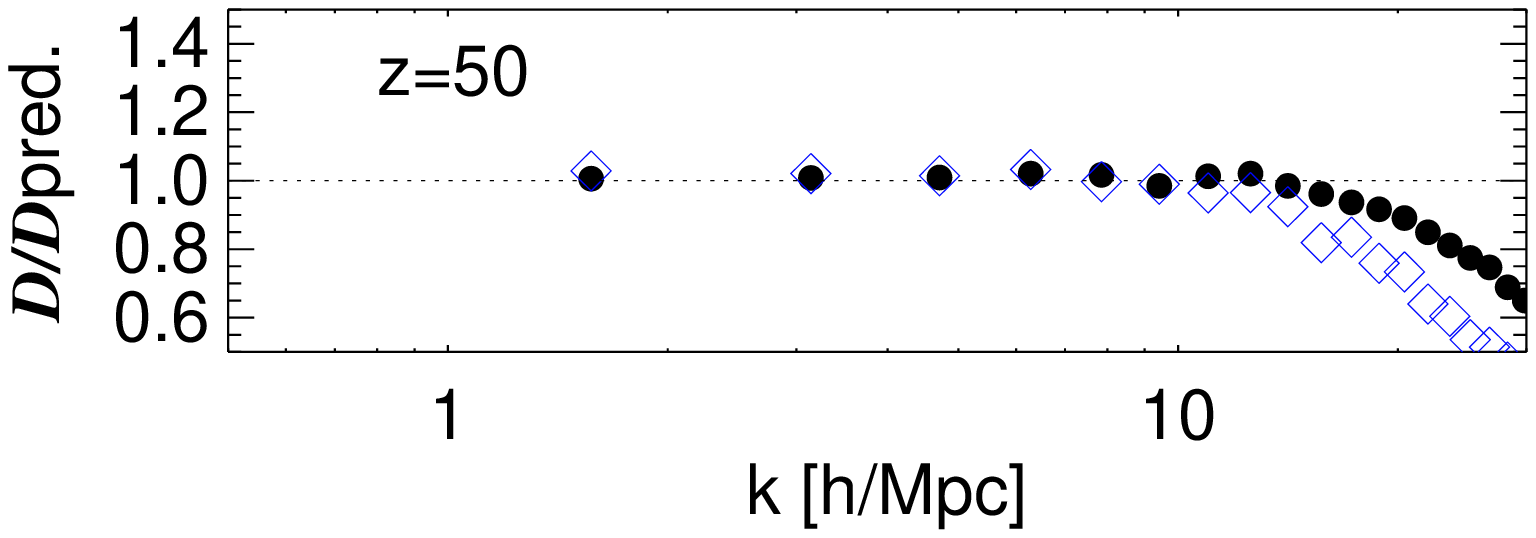}%
\epsfxsize=0.45\hsize\epsffile{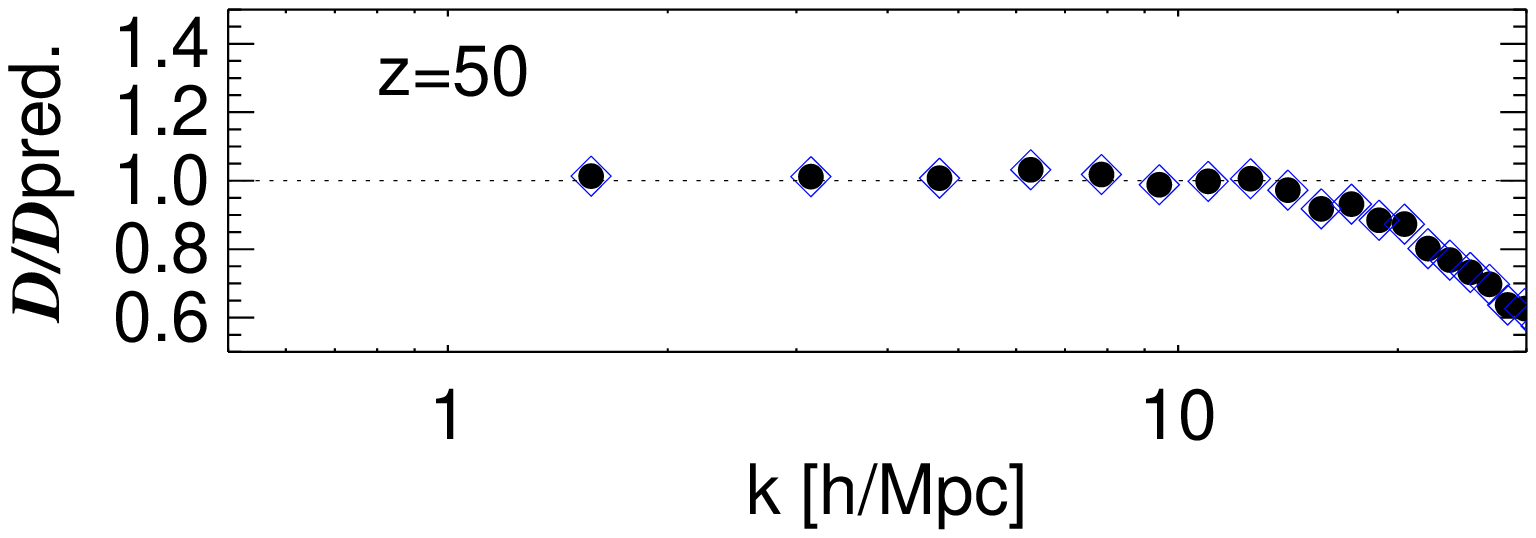}\\  
\caption{The measured power spectra for the baryonic and CDM
components are compared with the theoretical predictions 
at the initial epoch ($z=100$) and at $z=50$.
In each panel the top portion shows the power spectra $4 \pi k^3 P(k)$
and the bottom portion shows the measured growth rate normalised
by the theoretically expected value. Note the vertical axis in the
bottom portion is in linear scale.}
\end{figure*}

\begin{figure*}
\epsfxsize=0.45\hsize\epsffile{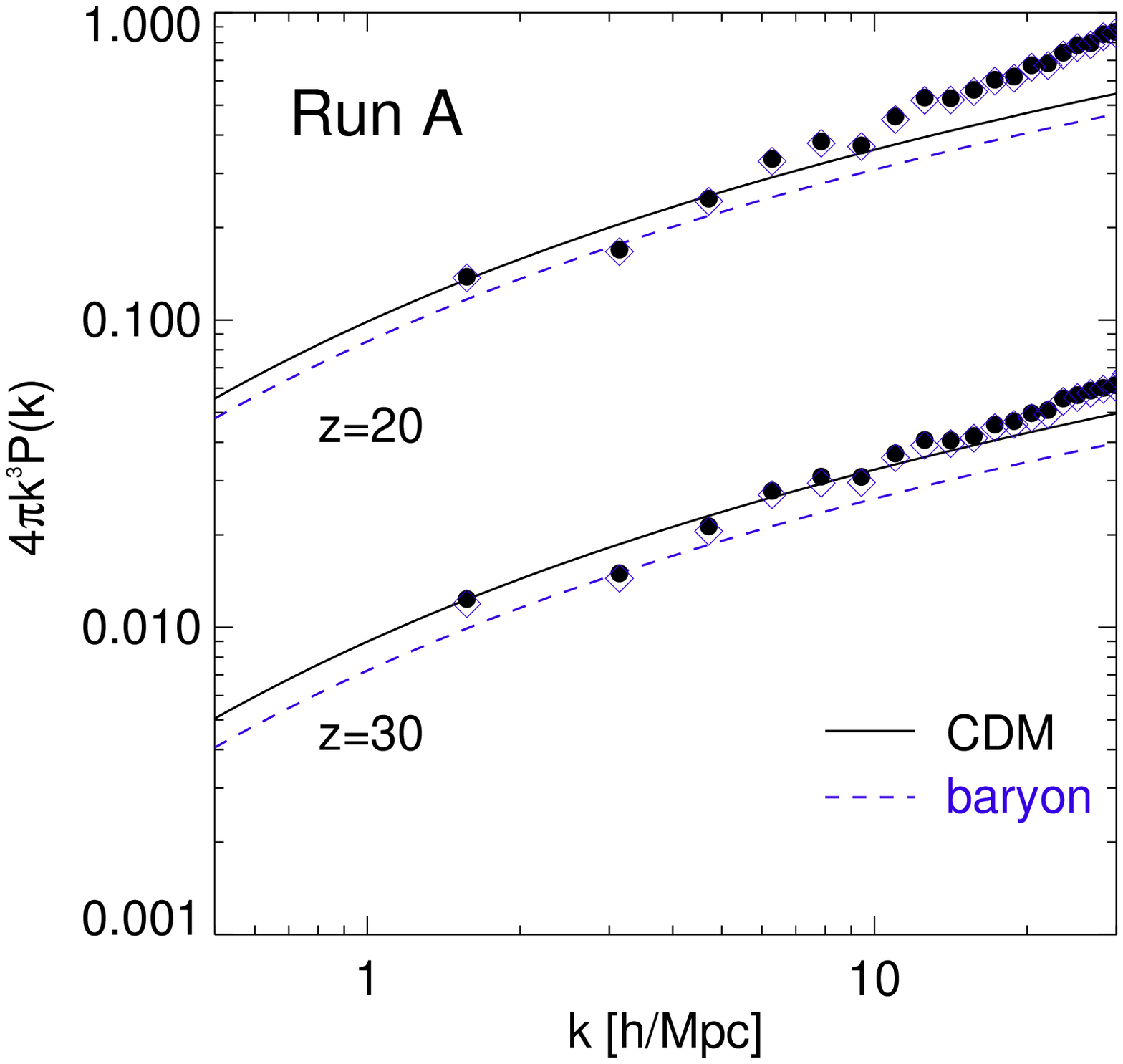}%
\epsfxsize=0.45\hsize\epsffile{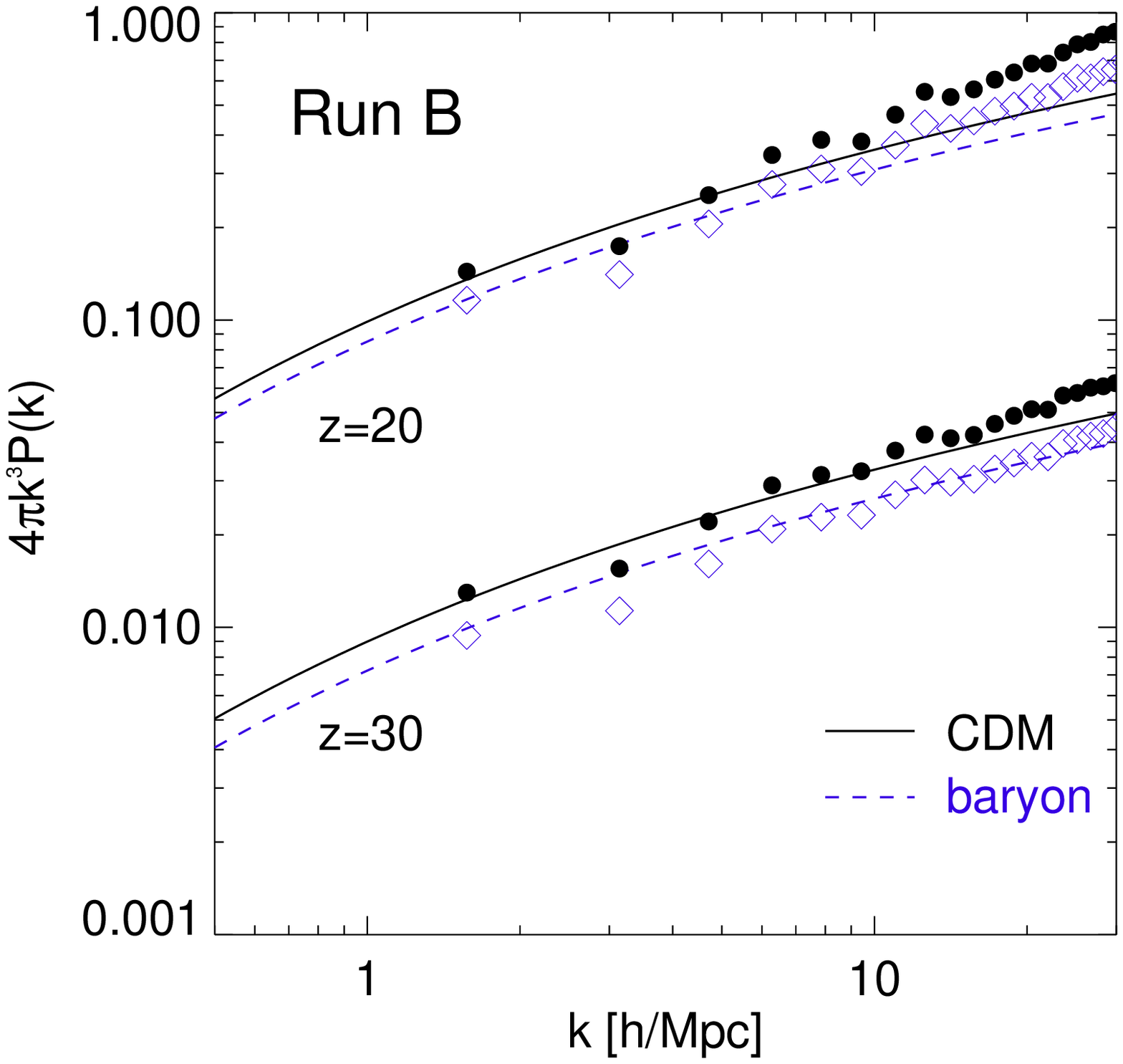}\\ 
\epsfxsize=0.45\hsize\epsffile{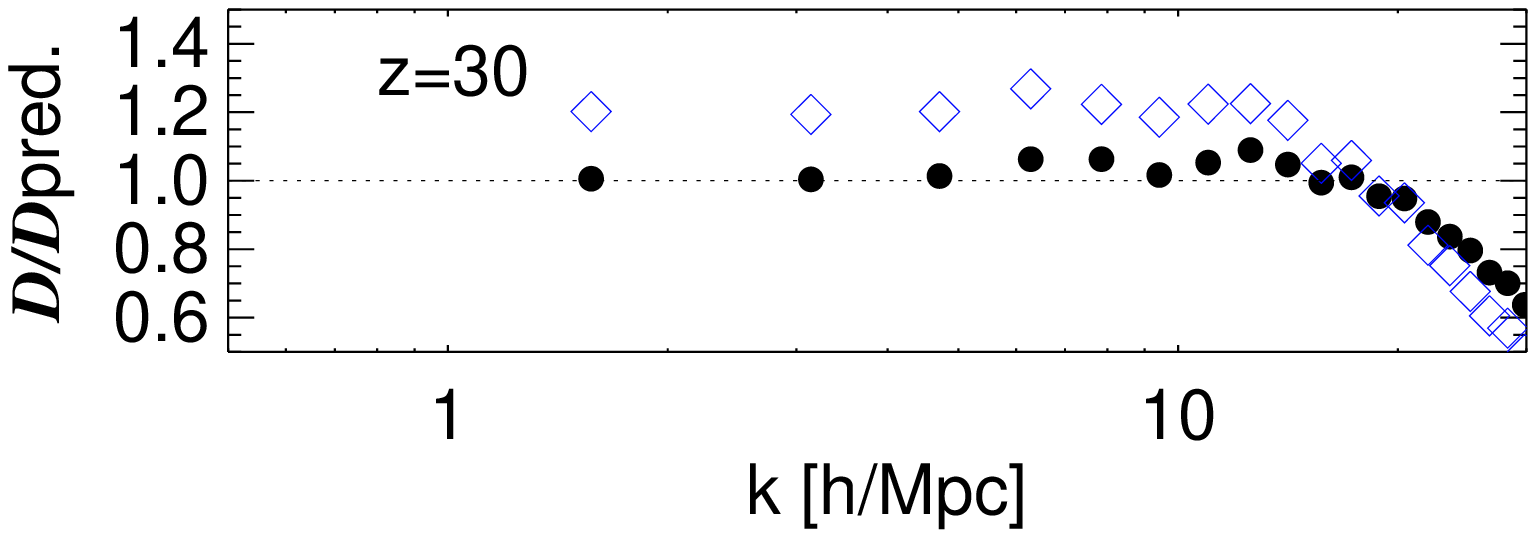}%
\epsfxsize=0.45\hsize\epsffile{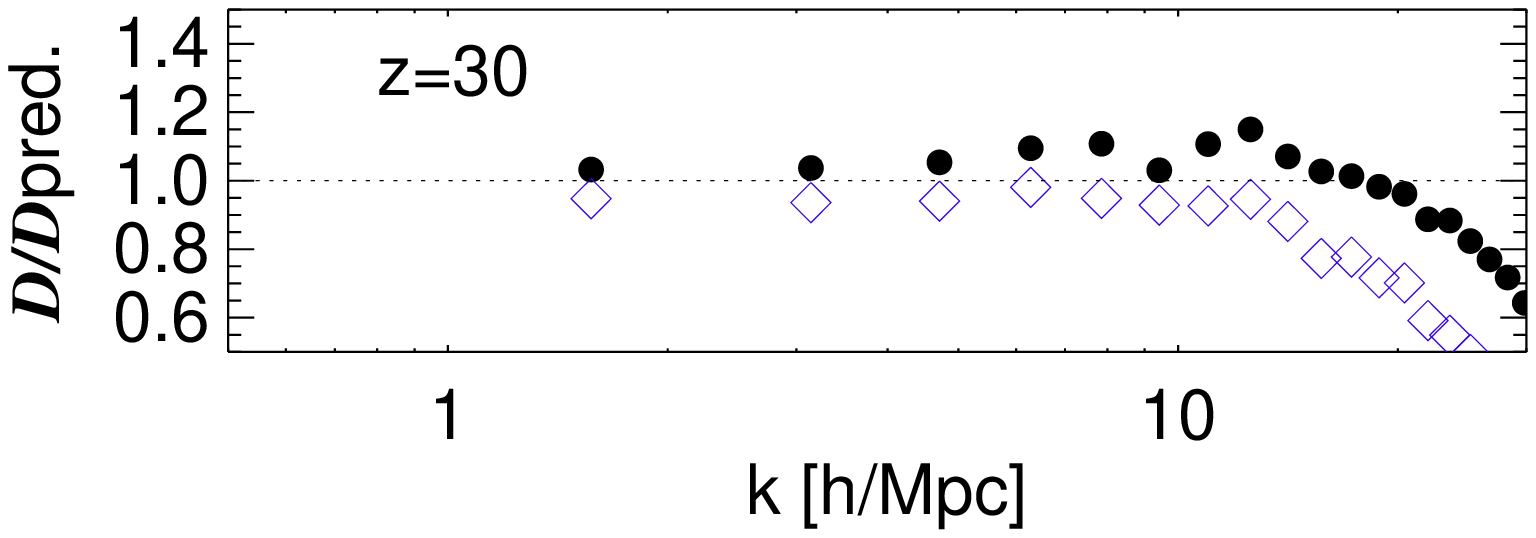}\\  
\epsfxsize=0.45\hsize\epsffile{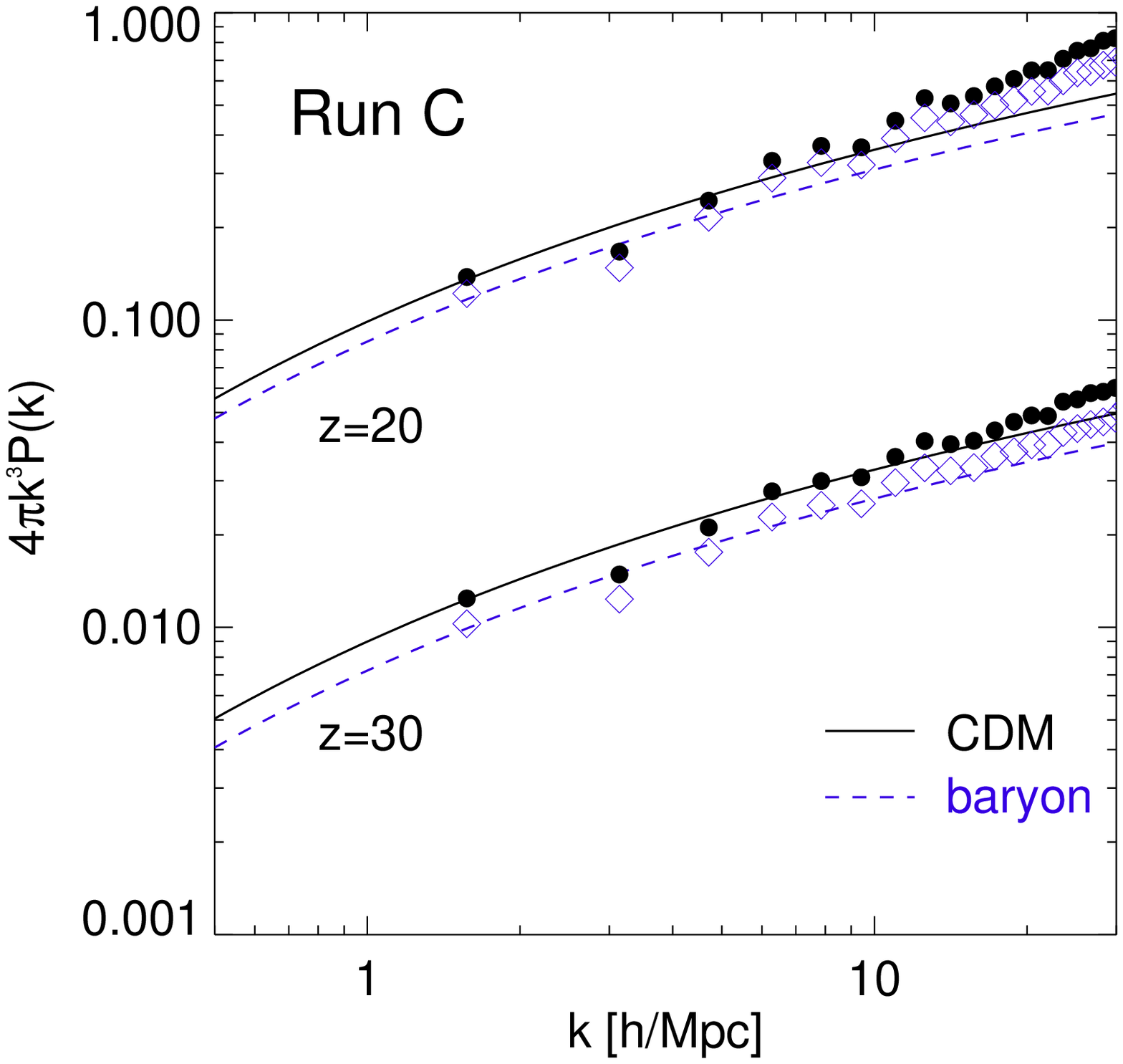}%
\epsfxsize=0.45\hsize\epsffile{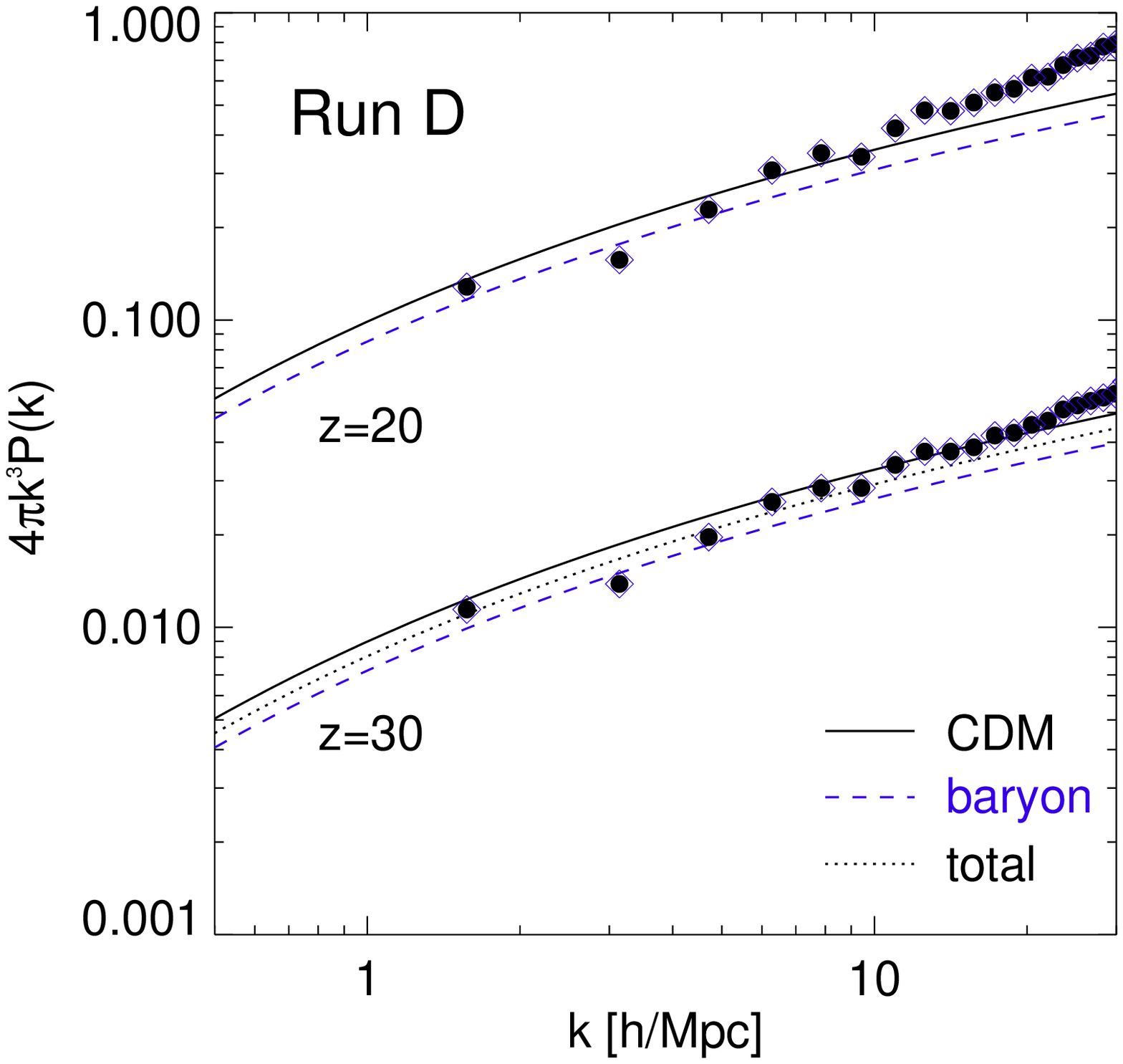}\\  
\epsfxsize=0.45\hsize\epsffile{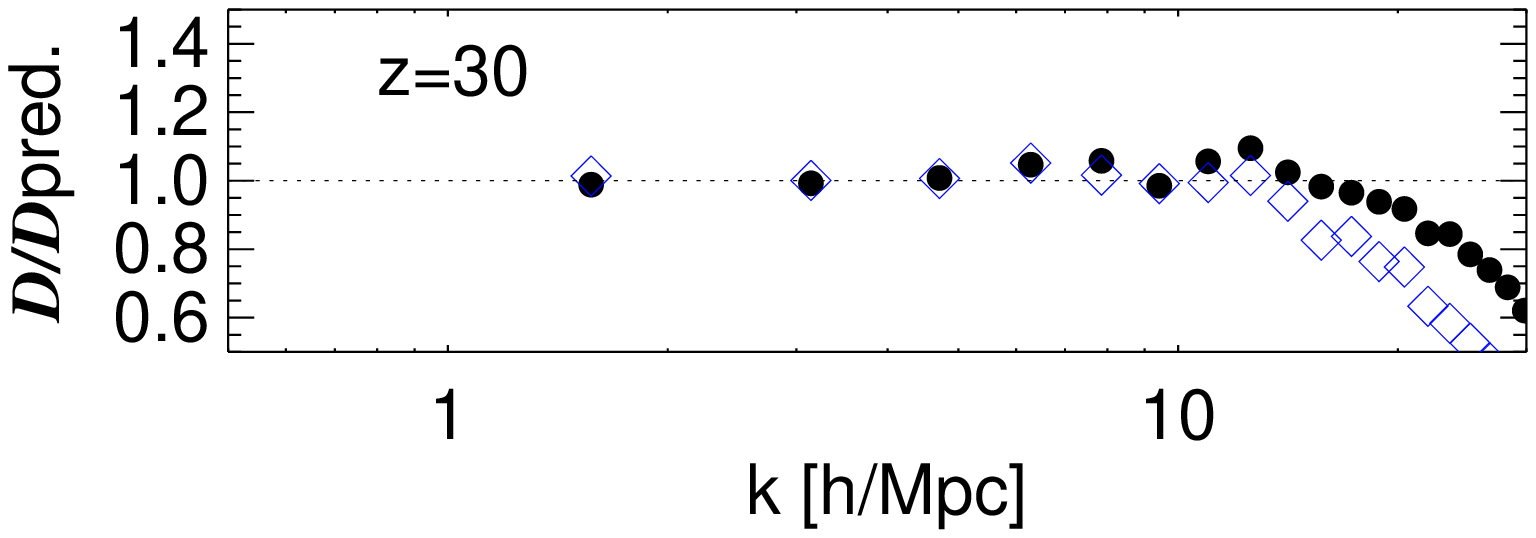}%
\epsfxsize=0.45\hsize\epsffile{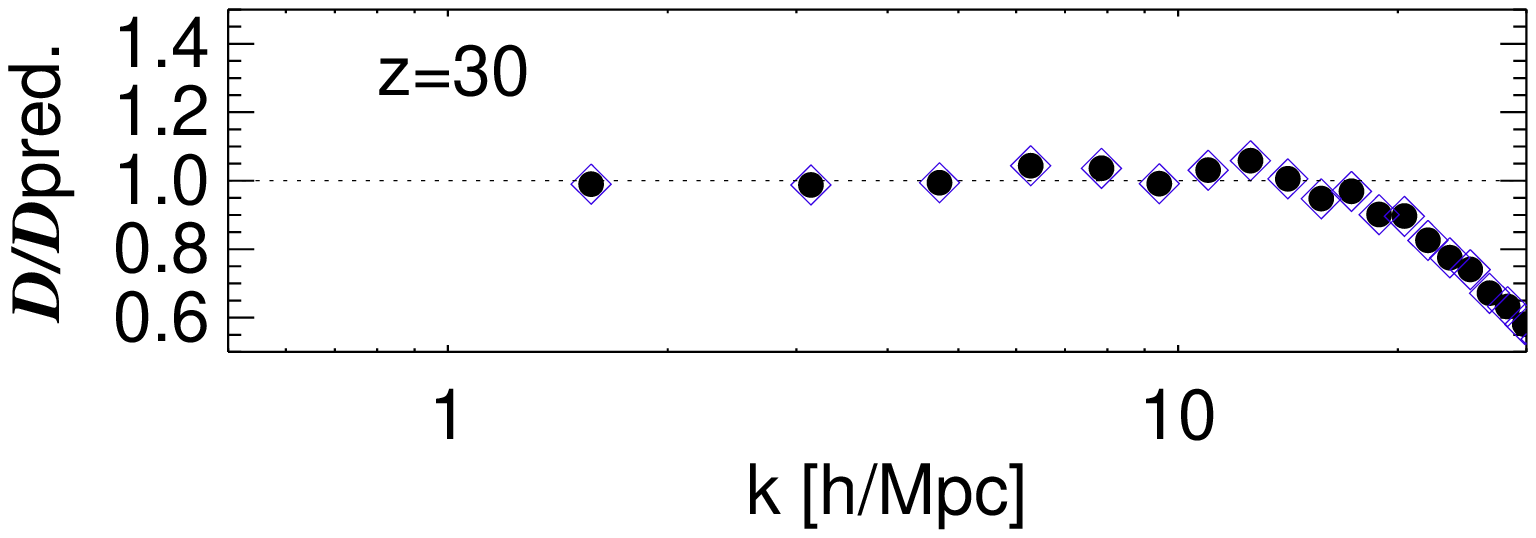}\\  
\caption{As for Figure 2, but for $z=20$ and $z=30$. 
The power spectra at $z=20$ are scaled 
vertically by a constant factor of five, in order to
clarify the plot}
\end{figure*}

\section{Results}

Figure 2 shows the power spectra of the initial mass distribution for
the four simulations (Run A, B, C, D), and the subsequent evolution is
shown in Figure 2 (for $z=50$) and Figure 3 ($z=20, 30$). 
The filled circles are measured power spectrum for CDM and the open diamonds are
for baryons. Note that the particular random realization for these runs
shows a dip at $k\sim 3 h$/Mpc. This is simply due to randomization 
when assigning the fluctuation amplitudes for each perturbation mode,
and irrelevant to the analysis in the present paper.      
Theoretical predictions are shown by the solid line
(CDM) and by the dashed line (baryon) for each output redshift.  The
measured power spectra at $z=100$ show a steep turn-over at large
wavenumbers $k > 20$ due to the Poisson noise which scales with the
particle number $N$ as $\left<|\delta_{k}|^2\right>=1/N$.  (Note that
we plot $4\pi k^3 P(k)$ in Figures 2 and 3.)  Although the level of
the Poisson noise could be reduced by using a larger number of
particles, we refrain from carrying out such large, costly
simulations. Since we focus on density fluctuations on large length
scales, $k < 20$, where the power spectra and the growth rates can be
measured robustly, the shot noise effect is not important in the
analysis presented in the following.

In Run D, baryons and CDM have exactly the same velocities both in
amplitude and direction. For this simulation the gas particles are
simply put on top of the CDM particles and thus they move together
initially.  Since we used the transfer function for the total matter
density in Run D, the measured power spectrum for CDM is smaller than
the theoretical prediction (solid line), whereas that for the baryons
is larger than the theoretical prediction (dashed line).  These
differences can be easily accounted for by the ratio of $\Omega_{\rm
baryon}$ and $\Omega_{\rm cdm}$ to $\Omega_{\rm total}$.

In the bottom portion of each panel in Figure 2, 3, we plot the growth
rate of the densities.  We measure the growth rate between $z_{\rm
init}(=100)$ and $z$, normalised by theoretical prediction as
\begin{equation}
\frac{D}{D_{\rm pred}} = \frac{P(k;z)/P(k;z_{\rm init})}{P_{\rm pred}(k;z)/P_{\rm pred}(k;z_{\rm init})}.
\end{equation}
We compute the theoretical prediction directly from the result 
of the Boltzmann solver.
Run C and Run D reproduce the theoretically predicted value very well,
whereas other runs show substantial deviations from the prediction.
This remains true even at $z=30$, as shown in Figure 3.

The significant over-shoot of the baryonic component in Run A is
explained by the tight coupling of the gas and CDM particles {\it
falsely caused} by using correct transfer functions.  In Figure
\ref{couple}, we plot the initial particle distributions in a slab of
thickness 40 $h^{-1}$kpc for Run A (top panel) and Run B (bottom
panel).  Figure \ref{couple} clearly shows that the gas particles are
placed too closely to the nearest CDM particles in Run A. Using
separate transfer functions whose amplitudes differ by an
approximately constant (and small) factor over the relevant length
scales, one obtains a distribution of gas particles each of which
is only slightly separated from the closest CDM particle.  In this
situation, a major fraction of the force exerted on a gas particle is
due to the nearest CDM particle, which causes the gas particle to move
faster than expected in linear theory. This false coupling was
successfully avoided in RunB, for which we used two sets of
independent glass distributions; one for the gas particles and the
other for the CDM particles.  Thus, the velocity vector of a gas
particle is uncorrelated with the direction to the nearest CDM
particle, as is seen in Figure \ref{couple}.

\setcounter{figure}{3}
\begin{figure}
\centering
\epsfxsize=0.9\hsize\epsffile{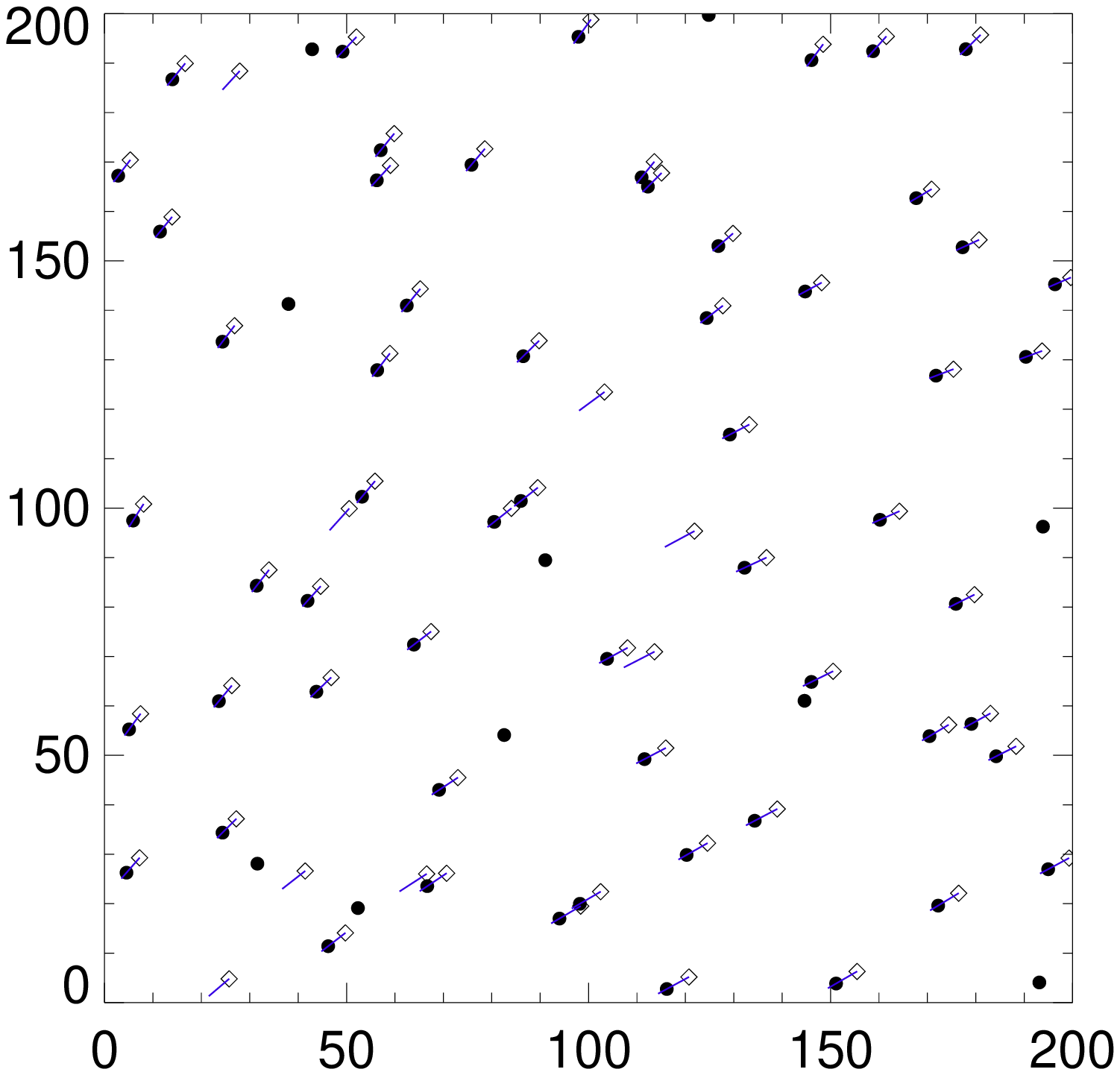}\\
\epsfxsize=0.9\hsize\epsffile{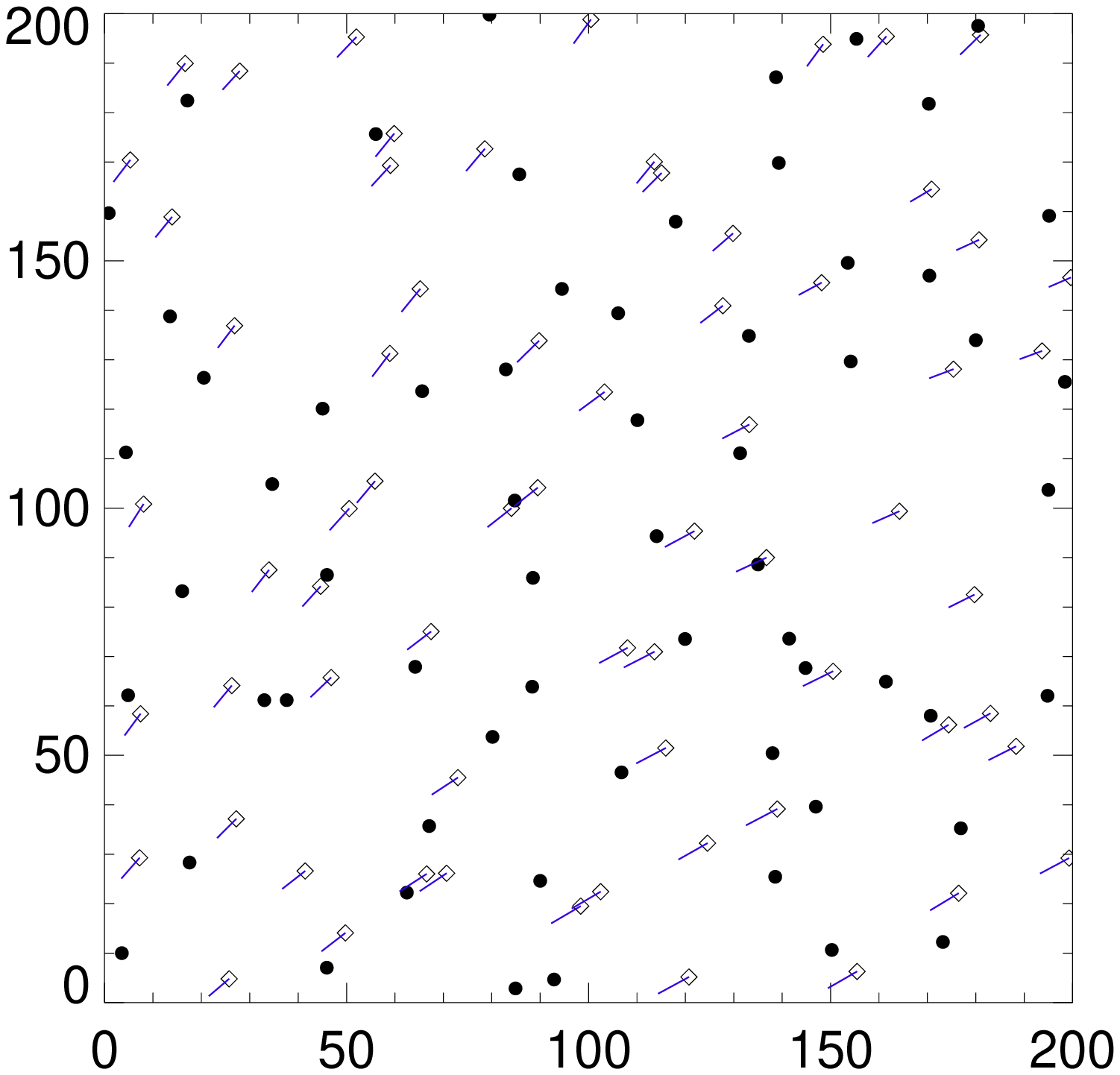}\\
\caption[Distribution]
{The initial particle distribution in a slab of thickness 40
$h^{-1}$kpc (1 percent of the box side-length) in Run A (top) and Run
B (bottom).  The solid points are CDM particles and the open diamonds
are gas particles. The short lines originating from the gas particles
show the velocity vectors. }
\label{couple}
\end{figure}

Although Run B appears to behave reasonably well, CDM density
fluctuations grow slightly too fast whereas the baryon density
fluctuations grow too slowly, as seen in the growth rate in comparison
with the theoretical prediction.  The initial velocities are assigned
separately to each component following the standard procedure of the
Zel'dovich approximation.  Without correcting velocities, the CDM
density fluctuations grow faster than the correct solution, and the
opposite is true for the baryons.  This feature is not seen in Run C,
because we corrected the initial velocities for both the baryonic and
CDM components.  Indeed, the density fluctuations in Run C grow almost
precisely as predicted from $z_{\rm init}=100$ to $z=20$.

Run D reproduces the correct growth of density fluctuations.  This is
as expected, because, in Run D, the two components initially behave as
a single component, until the pressure forces on gas particles becomes
important and, making the orbits of the gas and CDM particles deviate
from each other. Note that, at any output redshift, the density
fluctuations for a single component (baryons or CDM) does not
accurately represent the true solution.  Hence, we conclude that the
set-up for Run D is not suitable for studies that are concerned with
the difference in the distribution of baryons and dark matter.

Overall the initial set up for Run C is clearly the most favourable.
The method used for Run C appears to eliminate all the problems and
undesirable features found in the other simulations. Also, the
excellent agreement between the measured power spectra and the
theoretical prediction for {\it both} baryonic and CDM components over
a large redshift interval confirms that the method is useful for the
multi-component cosmological simulations.

\begin{figure}
\centering
\epsfxsize=\hsize\epsffile{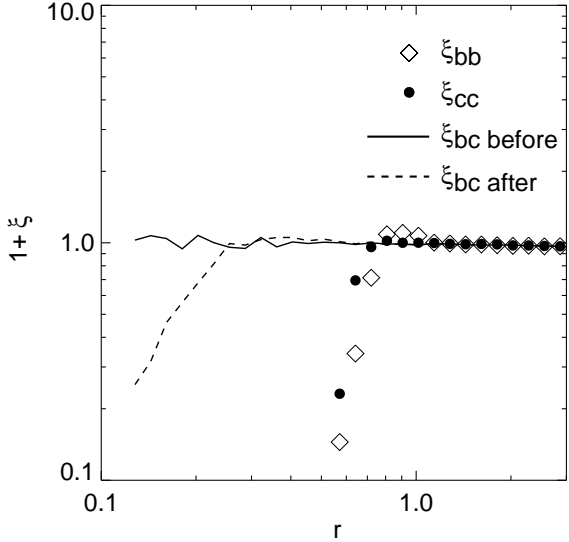}\\
\caption[Correlation]
{The two-point correlation function for the gas particles ($\xi_{\rm
bb}$, open diamonds) and for the dark matter particles ($\xi_{\rm cc}$,
solid circles) in a glass distribution. The solid line shows the
cross-correlation $\xi_{\rm bc}$ computed for an arbitrary superposition
of two independent glass files, whereas the dashed line is the
cross-correlation measured for the distribution of the mixture of the
two components after they were evolved for a brief period (see text). The separation
length $r$ is normalised by the mean inter-particle separation.}
\label{cross1}
\end{figure}

\begin{figure}
\centering
\epsfxsize=\hsize\epsffile{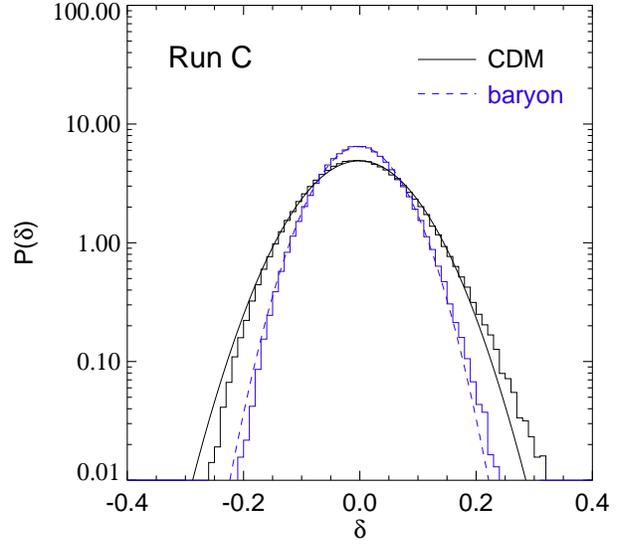}
\epsfxsize=\hsize\epsffile{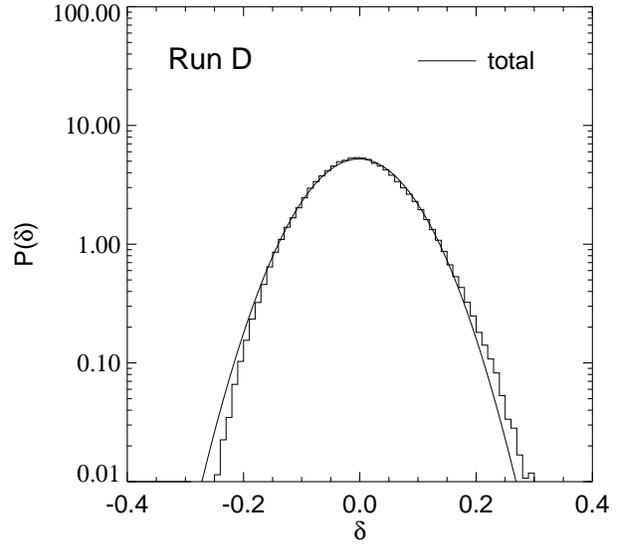}
\caption[Gaussian]
{The probablity distribution $P(\delta)$
for the initial conditions for Run C (top) and Run D (bottom).
In Run D, the CDM and gas particles have identical distibutions
and hence $P(\delta)$ is also identical initially.}
\label{gauss}
\end{figure}

\section{Discussion}

\subsection{Glass distribution for two components}
Although the comparison of the results of Run A and Run B show the
advantage of using two glass distributions for the CDM and baryonic
components, it remains unclear whether the mixture of two independent
glass distributions still possesses all desired features.  Generally
it does not, because close encounters between particles from the
different components
could happen (Jun Makino, private communication).  In order to address
this issue, we compute the cross-correlation function $\xi_{\rm bc}$
for a mixed glass distribution.  We make a distribution for $2\times
128^3$ particles by arbitrarily superposing a glass distribution for
$128^3$ particles on another glass distribution which is independently
generated.  The solid line in Figure \ref{cross1} shows the gas-dark
matter cross-correlation $\xi_{\rm bc}$ for this arbitrary mixture of
two sets of glass distributions. The overall behaviour might seem
quite good, as the cross-correlation $\xi_{\rm bc}$ stays at an almost
constant value around zero. However, it means that there is some
chance that a gas particle finds a CDM particle quite close to it.
Since this chance coupling of the two different components can cause
exactly the same problem, even locally, as we found in Run A, it is
preferable to generate a distribution of the two components such that
the cross-correlation between them is reduced on very small scales.
We follow the same procedure as we have done to create glass
distributions; i.e., we switch the sign of the gravitational force and
evolve the system consisting of a mixture of $2\times 128^3$
particles in an expanding background. Since our primary purpose is to
avoid false coupling of gas and dark matter particles, we evolve the
system only for a brief period of time.  Figure \ref{cross1} shows the
cross-correlation function for the initial (solid line) and
evolved (dashed line) mixture of the particles. We also plot the
correlation of the gas particles (open diamonds) and that of the CDM
particles (solid circles).  Figure \ref{cross1} shows that the glass
distribution we obtained in this manner possesses the following
desired properties: (1) the correlation of a single component is
reduced on scales below the mean-interparticle separation, and (2) the
cross-correlation is also reduced on the small scales, confirming that
the false coupling with the other component such as that found in Run
A will not be induced even when we use separate transfer functions.

\begin{figure}
\centering
\epsfxsize=\hsize\epsffile{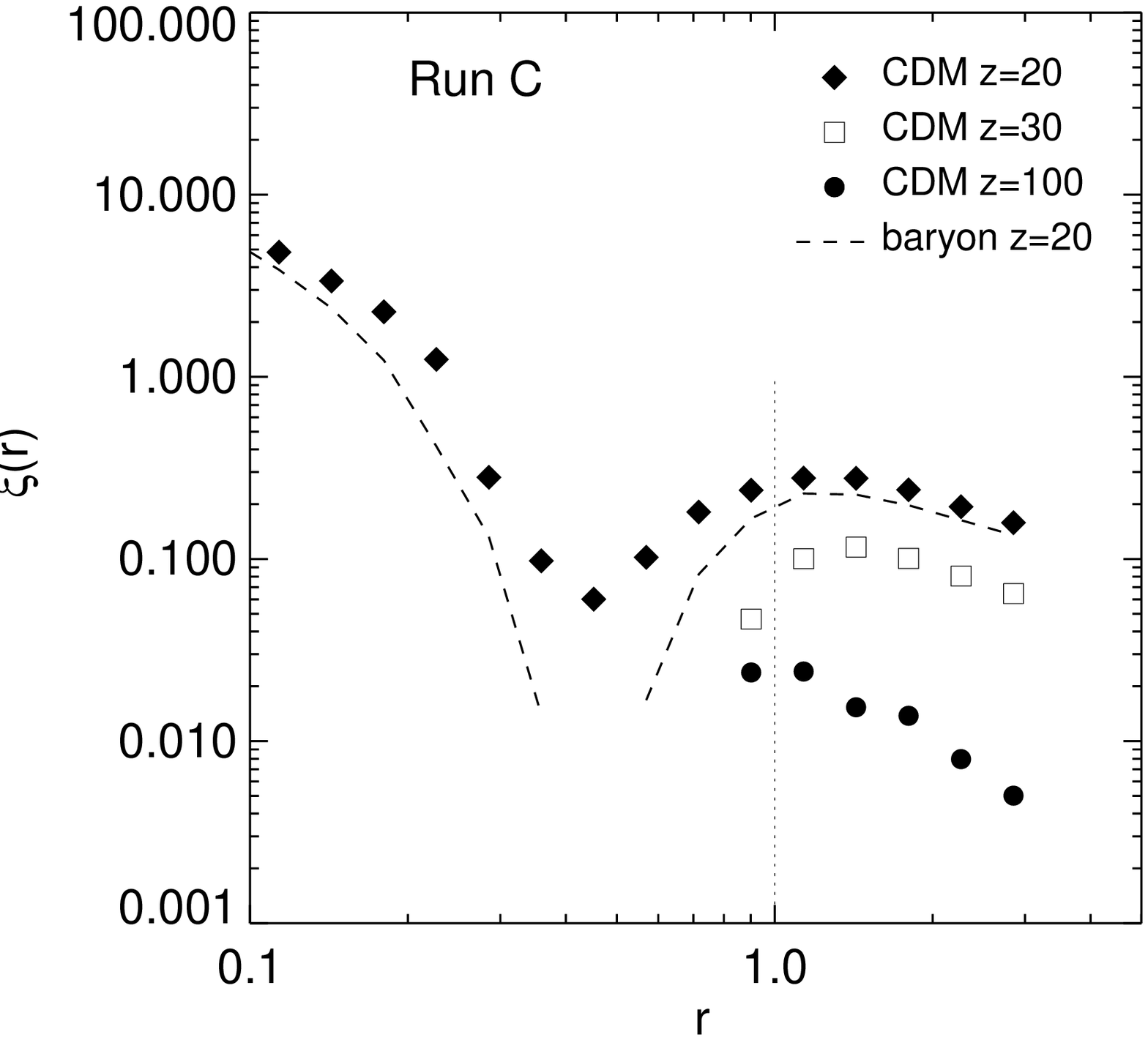}
\epsfxsize=\hsize\epsffile{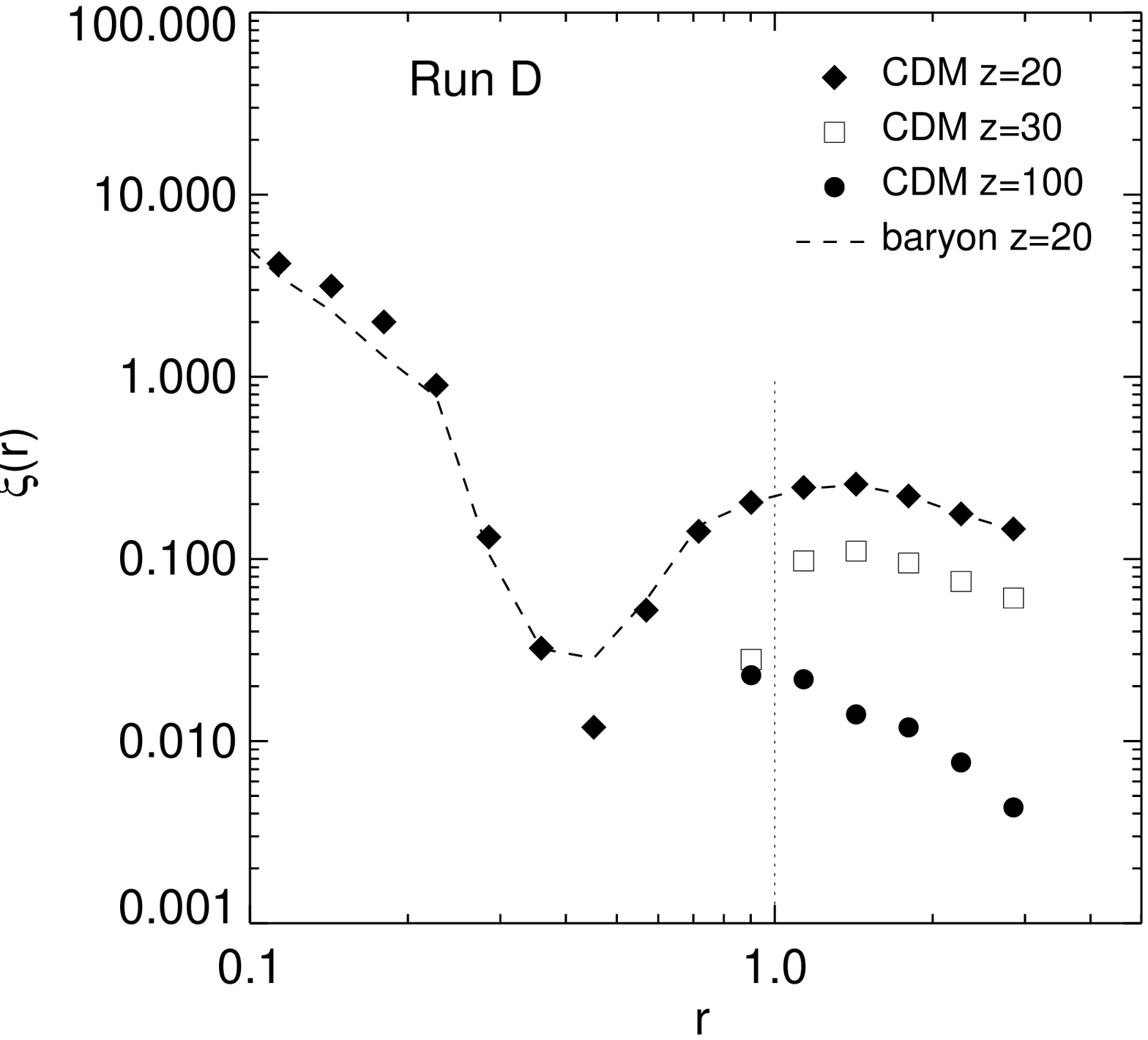}
\caption[Gaussian]
{The evolution of the two-point correlation function 
for Run C (top) and Run D (bottom). Symbols are for the dark matter component,
as indicated in the legend in each panel, and the dashed line
is the correlation function of the baryonic particles at $z=20$. The separation
length $r$ is normalized by the mean inter-particle separation.}
\label{2pcf_evol}
\end{figure}

\subsection{Real space correlation on small scales}
While we conclude that glass distributions are preferred 
over a regular grid distribution for the reasons described in the
previous sections, there still remain a few issues to be clarified 
regarding unperturbed particle distributions.
Setting-up initial conditions for cosmological particle 
simulations is generally divided into two parts. One is the generation of a uniform
particle distribution, which we have discussed, and the other is imposing desired density 
fluctuations by displacing the particles using the Zeldovich approximation (section 2.2).
For the currently standard cosmological models, we further assume that 
the initial density field is a Gaussian random field, for which the power
spectrum fully describes the statistical properties. 
We first check and show that the initial particle distribution generated in this manner
represents a Gaussian density field. i.e., the probability distribution $P(\delta)$ of
overdensity $\delta({\rm x})=\rho/\bar{\rho} -1$ in real space is Gaussian.
Figure \ref{gauss} shows $P(\delta)$ for our Run C and Run D.
We compute the real space density $\delta({\rm x})$ by first re-sampling
the particles onto a $128^3$ grid and then smoothing on a scale $R_{s}=100 h^{-1}$kpc.
For Run C, we compute $P(\delta)$ for both the baryonic and dark matter components,
whereas for Run D only the dark matter component is used.
(Note that in Run D the gas particles are put on top of the dark matter particles initially,
and hence $P(\delta)$ at the initial epoch is identical.)
As clearly seen in Figure \ref{gauss}, 
the computed probability distribution is well fitted by a Gaussian for each case.
The difference between the two
components in Run C is due to the difference in the linear power amplitude
on the chosen length scale $R_{s}\sim 100 h^{-1}$kpc. The measured variances
for the baryonic and dark matter components are $\sigma_{\rm b}$=0.062, 
$\sigma_{\rm cdm}$=0.081, respectively,
in good agreement with analytical estimates $\sigma_{\rm th, b}$=0.060, 
$\sigma_{\rm th, cdm}$=0.083 computed from the input density power spectra.
The corresponding numbers for Run D are $\sigma_{\rm total}$=0.076 and $\sigma_{\rm th, total}$=0.078.
We note that, in practice, we used a random number generator in assigning the phases for each Fourier mode
so that the generated initial conditions possess a desired feature of a Gaussian random field. 

Since any particle distribution (e.g. glass or grid) has its 
own intrinsic correlations (Gabrielli et al. 2002), the perturbed distribution will have, 
in principle, a superposition of the intrinsic and imposed (desired) correlations. 
Baertschiger \& Sylos-Labini (2002) argue that
the desired power law correlations for cosmological models are not properly 
realized in discritised density fields at the initial epoch (see also
Knebe \& Dominguez 2003).
We quantify and show this feature by computing the real space correlation
from particle distributions.
We use the pair-count estimator proposed by Landy \& Szalay (1993):
\begin{equation}
\xi (r) = \frac{DD(r)-2DR(r)+RR(r)}{RR(r)}.
\end{equation}
We randomly distribute the same number of particles as the actual simulation particle,
and evaluate the data-data (DD), data-random (DR), and random-random (RR)
terms to obtain $\xi (r)$.
In Figure \ref{2pcf_evol} we show the two-point correlation functions
measured from the outputs of Run C and Run D.
Initially the correlation on scales below the mean inter-particle separation (dotted line)
is absent, reflecting the damped correlation on the small scales in 
the unperturbed distribution (see Figure \ref{cross1}).
The particle clustering at early epochs 
on the smallest scales is considerably affected by the discreteness as
argued by Hamana, Yoshida \& Suto (2002).
Baertschiger et al. (2002) claim that the development of a power-law correlation 
on the small scales is governed partly by the initial clustering. 
It is interesting that the two-point correlation functions of the two components
at $z=20$ are quite similar even on the smallest length scales, 
despite the fact that they started from {\it different} glass distributions.
On the other hand, the feature that the correlation amplitudes are different between 
the two components in Run C is plausible, and it appears to reproduce correctly
the smaller density fluctuations of baryons.
At late epochs the nonlinear power transfer 
eventually dominates the initial power if the effective spectral index at the scale
$n\equiv d\log P(k)/d\log k$ is much less than -1. CDM models
satisfy this condition on small length scales (Suto \& Sasaki 1991), 
and hence the power on scales below the mean
particle separation at later epochs is expected to be generated via nonlinear mode-coupling
(Suginohara et al. 1991; Hamana et al. 2002). 
Further theoretical study is clearly needed to understand the behaviour of 
the small scale correlations at early epochs.

\subsection{Differing phases}

\begin{figure}
\centering
\epsfxsize=\hsize\epsffile{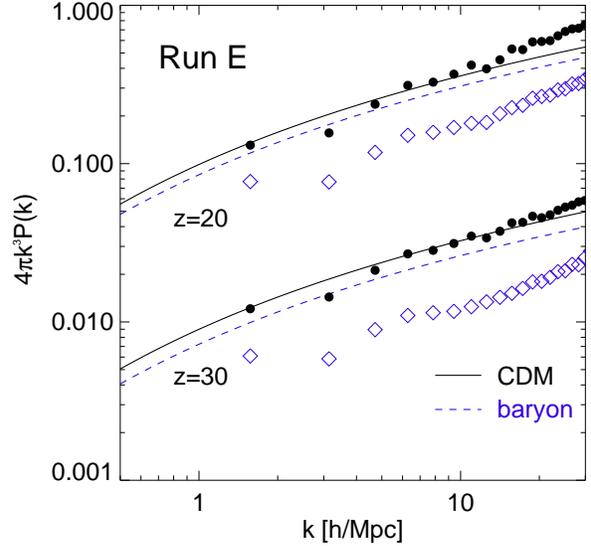}
\caption[Distribution]
{The evolution of the power spectra in Run E.
As in Figure 3, the power spectra at $z=20$ are scaled 
vertically by a constant factor of five, in order to
clarify the plot.}
\label{phase}
\end{figure}

As we discussed in Section 3.3, the phases of the perturbations are
assumed to be identical for baryons and dark matter as long as the
density perturbations are in the linear regime.  Nevertheless, it is
interesting to examine how the growth of density fluctuations is
affected by offsets of phases between the two components. 
We note that, in real astrophysical situations, phase offsets may be caused 
during early reionization when the intergalactic gas is photo-heated 
by radiation from an early generation of stars and driven out of dark matter 
potential wells.
In such cases, it is expected that the growth of either or both components 
is delayed, compared to the case
where the phases are synchronised.  We carry out a simple test by
applying independent phases to baryon and CDM perturbations at the
initial epoch.  For simplicity, we assign completely random phases,
rather than specifying actual mechanisms that could cause such an
offset.  Hence, the test is meant to show the effects of the phase
offset qualitatively. 
Figure \ref{phase} shows the evolution of the density
fluctuations for Run E.  As expected, the evolution of both the baryon
and the CDM density fluctuations is delayed, with the effect on the
former being more substantial.  On the other hand, it is interesting
that the growth of the baryon density fluctuations is not entirely
prevented, despite the large difference in the fractional densities
between the two components.  This is because, on large scales, the
dark matter density fluctuations are still small and the gravitational
force from the dark matter has not significantly changed the initial
momentum of the baryonic components.
Note that, in Figure \ref{phase}, the apparent differences between the measured
power spectra and the theoretical prediction should be interpreted to
be owing mostly to the difference in the background assumption,
because the theoretical prediction itself is obtained assuming the
initial phases are identical.

\subsection{Nonlinear objects}

\begin{figure}
\centering
\epsfxsize=0.9\hsize\epsffile{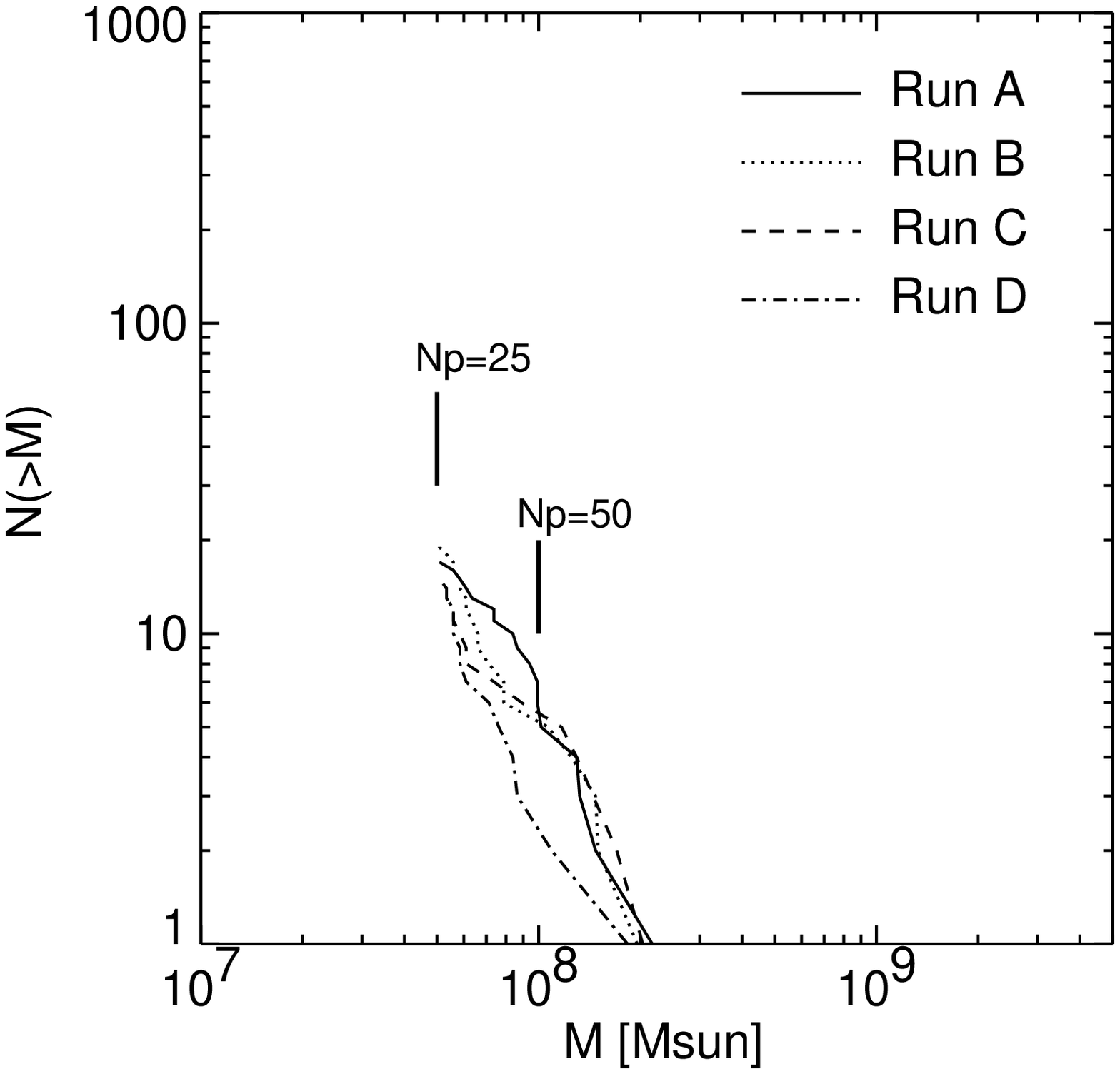}
\epsfxsize=0.9\hsize\epsffile{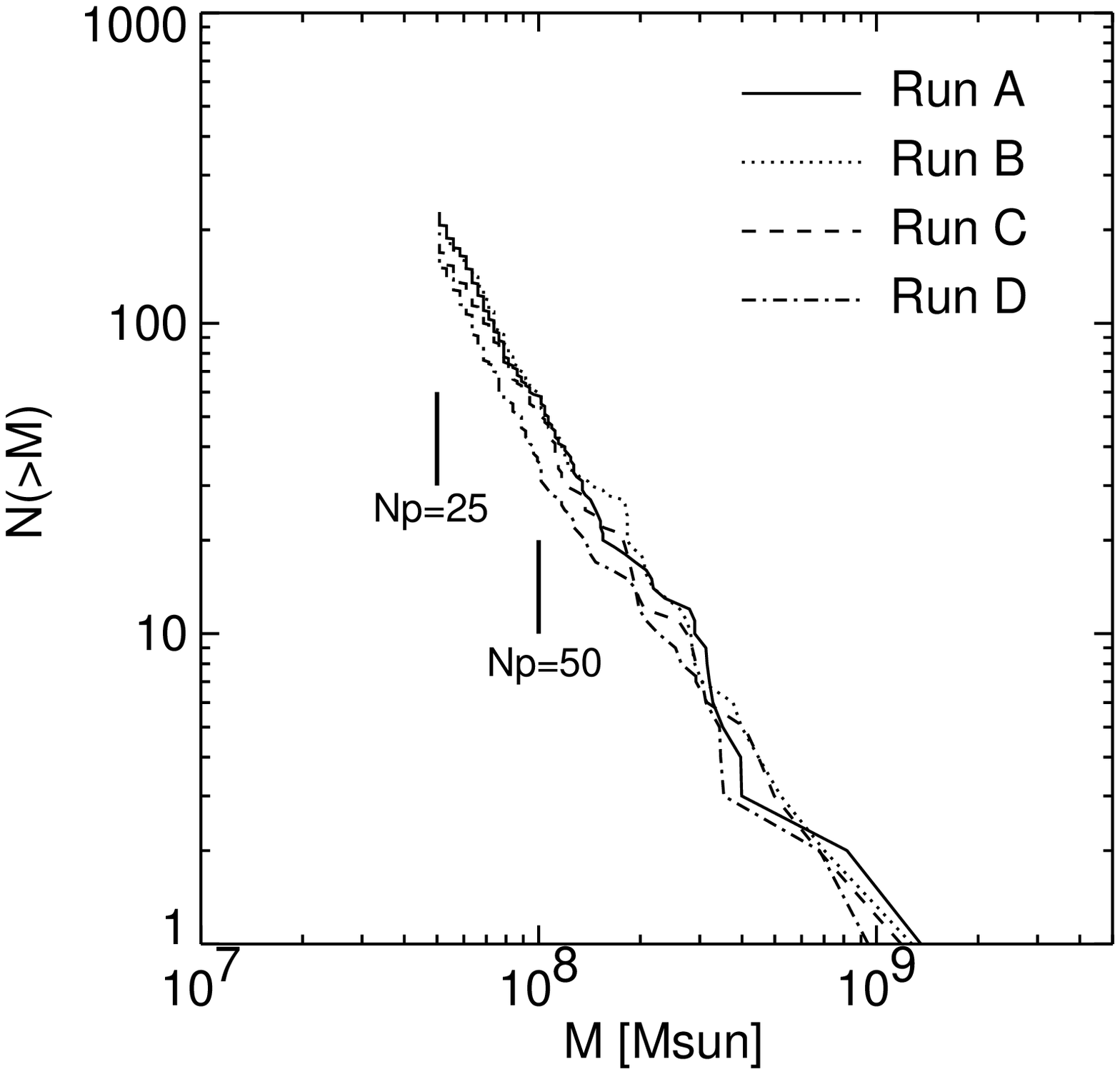}
\caption[Distribution]
{Cumulative mass function of dark matter halos
at $z=20$ (top) and $z=15$ (bottom). The vertical lines indicate
the number of particles constituting a halo of a corresponding mass.}
\label{mf}
\end{figure}

In the previous sections, we have focused on the evolution of density
fluctuations on large scales when they are in linear regime.  As
Figure 2 and Figure 3 show, the overall differences between our
simulations appear as if their effective growth factors deviate,
falsely or due to an inaccurate set-up, from the {\it true} value.
This raises an interesting question of whether or not such differences
also show up in some properties of non-linear objects.  Using our
simulations, we identify dark matter halos by running a friend of
friends (FOF) group-finder with the standard choice of linking parameter
$b=0.164$.  We first consider only the dark matter component. The
baryon fraction within the halos will be examined later in this
section.  Figure \ref{mf} shows the cumulative mass functions for our
simulations at $z=20$ and $z=15$. Since the overall abundance of halos
is still small at $z=20$, as seen in the top panel in Figure \ref{mf},
we allow our simulations to evolve to $z=15$ to have a large number of
halos. We show the result for the $z=15$ outputs in the bottom panel
of Figure \ref{mf}.  We compute the mass of the identified halos to be
the number of member particles times the particle mass.  We discard
halos consisting of fewer than 25 particles, and thus the minimum mass
in our sample is set to be $5\times 10^7 h^{-1}M_{\odot}$.  Although
the mass functions measured in the four simulations agree reasonably
well, there is an appreciable differences, particularly between Run D
and the other runs. This is easily accounted for by the fact that the
growth of the dark matter density fluctuations in Run D is slower than
the other runs (see Figure 2, 3).

\begin{figure}
\centering
\epsfxsize=\hsize\epsffile{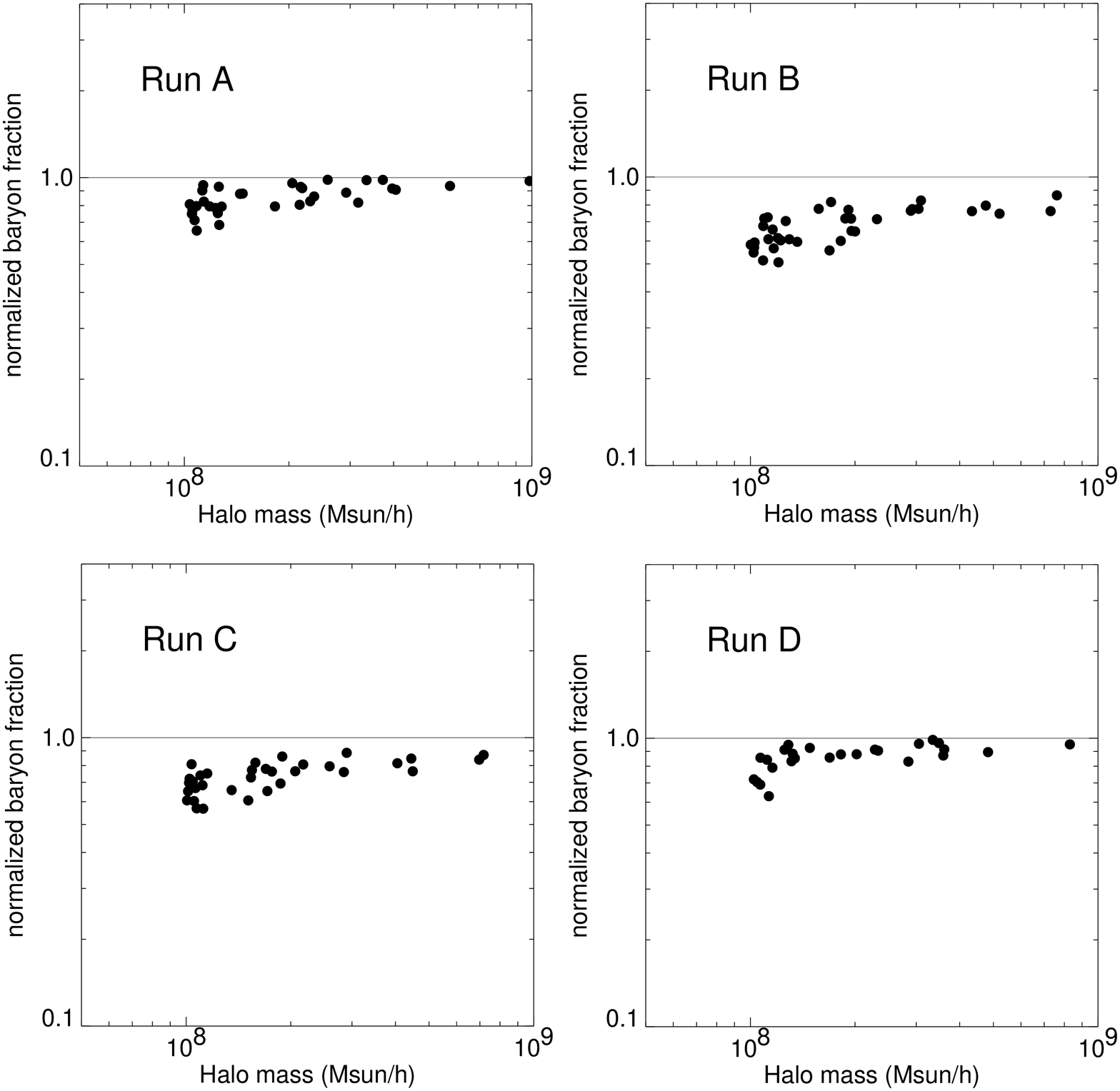}
\caption[Distribution]
{Baryon fractions for massive ($M>10^8 h^{-1}M_{\odot}$) 
halos at $z=15$. The baryon fractions are normalised by the global baryon 
fraction $\Omega_{\rm baryon}/\Omega_{\rm matter}$.}
\label{fb}
\end{figure}

We also measured the baryon fraction within the dark halos.  To this
end, we compute the virial radii and virial masses for the halos
following Yoshida et al. (2002).
Briefly, we define the virial radius $R_{\rm vir}$ as the radius of
the sphere centred on the most bound particle of the FOF group having
overdensity 200 with respect to the critical density. The virial mass
$M_{\rm vir}$ is then the enclosed mass (gas and dark matter) within
$R_{\rm vir}$.  We compute the baryon fractions for the halos as
$f_{\rm b}=M_{\rm gas}/M_{\rm total}$, and normalise by the global
baryon fraction $\Omega_{\rm b}/\Omega_{\rm matter}$.  Figure \ref{fb}
shows the measured baryon fractions for the halos in our simulation at
$z=15$. In all cases the (normalised) baryon fractions are smaller
than unity, representing correctly the effect of hydrodynamic pressure
in the absence of gas cooling.  Whereas the large scatter at masses
smaller than $\sim 2\times 10^8 h^{-1}M_{\odot}$, consisting of fewer
than 100 particles, is due to the mass resolution (see Yoshida
et al. 2002 for a discussion of the resolution issue), the difference
on larger mass scale between the four runs is unexpected.  Again, by
noting the residual difference in the baryon and dark matter density
fluctuations shown in Figure 3 and Figure \ref{2pcf_evol}, 
we are led to the conclusion that the
difference in the density fluctuations at early epochs is responsible
for the overall trend shown in Figure \ref{fb}.  Indeed, the
amplitudes of the baryon and dark matter density fluctuations in Run A
and Run D are quite similar from early on ($z\sim 50$), while those in
Run B and Run C remain relatively large until low redshift ($z\sim
20$).  Thus the gas infall is slightly delayed in Run B and Run C,
resulting in the smaller baryon fractions than in Run A and Run D.

It should be noted that these results are not expected to be precise
in details, but are likely to be dependent on the procedure for 
identifying halos and also on the definition of the halo mass.  
The detailed description
of nonlinear evolution is beyond the scope of our paper and thus we
reserve this issue for the subject of future work using simulations
with substantially higher resolution than those described here.

\section{Conclusion}

We have studied the evolution of the matter density fluctuations in
cosmological $N$-body/hydrodynamic simulations, and discussed various
aspects of the set-up of the initial conditions.  All the methods we
used in generating the initial conditions except for Run E may appear
to be {\it plausible}, differing only in a few details.  However, we
have found that such seemingly minor details cause appreciable
differences in the evolved density field, not only in linear regime
but also in the non-linear regime.

Interestingly, it has turned out that neither of the methods used for
Run A and Run D provide suitable initial conditions, with the results
of the former being more significantly affected by numerical
artifacts.  This perhaps comes as a surprise, particularly in the case
of Run A, because one may naively hope that using separate transfer
functions for baryons and CDM would produce more accurate results. In
fact, the baryon density fluctuations reproduced in Run A are the
least accurate in terms of the growth rate. Although the particular problem of
false particle coupling found in Run A is likely to be a problem
in simulations where both components are represented by discrete mass
elements, it is preferable, even in simulations that employ a grid-based
Eulerian scheme, to avoid artificial coherence between the initial fluid
velocity vector and the direction to the nearest CDM particle. 
This may deserve further study using a grid-based scheme.

By analysing the outputs of the numerical experiments, we identified
some peculiar problems and inaccuracies caused by the simplest
methods.  Then, based on the results, we have suggested a more
accurate method to set up initial conditions; i.e., the one used for
Run C.  In summary, we recommend: (1) using independent glass particle
distributions from which a ``glass mixture'' is made by a treatment to
avoid false coupling, (2) using two transfer functions computed for
baryons and for dark matter, {\it and} (3) taking into account the
difference in their velocities at the initialisation epoch.  
We emphasize that the latter two procedures may be
necessary for all cosmological simulations employing multiple
components, whether or not both dark matter and gas are realised with
particles.  We have explicitly shown that, in this manner, the
evolution of matter density fluctuations are accurately followed at
high redshift.

\section*{Acknowledgments}

We thank Edmund Bertschinger, Chung-Pei Ma, Matias Zaldarriaga, Hugues Mathis and
Takashi Hamana for valuable discussion. We also thank the anomnymous referee
for giving constructive comments which improved the manuscript. 
This work was supported in
part by NSF grants ACI 96-19019, AST 98-02568, AST 99-00877, and AST
00-71019, and by NASA ATP grant NAG5-12150.  The simulations were
performed at the Center for Parallel Astrophysical Computing at the
Harvard-Smithsonian Center for Astrophysics.

\end{document}